\documentclass[preprint,pre,aps]{revtex4}
\usepackage[dvips]{graphicx}
\begin{document}
\title{Density functional theory for soft matter with mesoscopic length scale fluctuations included within field-theoretic formalism}
\author{ A. Ciach}
\address{Institute of Physical Chemistry,
 Polish Academy of Sciences, 01-224 Warszawa, Poland}
 \date{\today} 

\begin{abstract}
 Mesoscopic theory for soft-matter systems that combines density functional and statistical field theory is derived from the microscopic theory by a systematic coarse-graining procedure. Grand-thermodynamic potential functional  for hard spherical particles that interact with (solvent mediated) spherically-symmetric potentials of arbitrary form is a sum of two terms. In the first term microscopic length-scale fluctuations are included, and the second term is the contribution associated with mesoscopic length-scale fluctuations. In the approximate theory the first term has the form of the density functional in the local density approximation, whereas the second term has the form known from the field theory and depends on the pair correlation function for which a pair of equations similar to the Ornstein-Zernicke equation with a particular closure is obtained. For weak ordering the theory can be reduced to the Brazovskii field theory with the effective Hamiltonian having the form of the grand-potential functional in the local density approximation. 

Within the framework of this theory we obtain and discuss the $\lambda$-line and the universal sequence of phases: disordered, bcc, hexagonal, lamellar, inverted hexagonal, inverted bcc, disordered, for increasing density well below the close-packing density. The sequence of phases agrees with experimental observations  and with simulations of many self-assembling systems. In addition to the above phases, more complex phases may appear depending on the interaction potentials. For a particular form of the short-range attraction long-range repulsion potential we find the bicontinuous gyroid phase (Ia3d symmetry) that may be related to a network forming cluster of colloids in a mixture of colloids and nonadsorbing polymers.
\end{abstract}

\maketitle
\section{Introduction}

Various soft-matter systems, including surfactant solutions, globular proteins, colloids in different solvents,  colloid-polymer or star polymer - linear polymer  mixtures, exhibit self-assembly into various  structures. In particular, clusters or micelles of various shape and size, including branched networks can be formed. The clusters or micelles may exhibit ordering into different periodic phases for a range of volume fractions, including quite dilute systems \cite{seul:95:0,gelbart:99:0,ciach:01:2,stradner:04:0,sedgwick:04:0,sanchez:05:0,campbell:05:0,stiakakis:05:0,andreev:07:0}. Phase-separation competes in such systems with formation of lyotropic liquid crystalline phases. Transitions between soft crystalline phases having different symmetries,  phenomena such as re-entrant melting\cite{arora:98:0,konishi:98:0,royall:06:0,ise:99:0}, and finally formation of hard crystals are observed when the volume fraction increases.
Theoretical studies of such systems can be based on liquid theories such as SCOZA, Landau-type theories or density functional theories (DFT). Unfortunately, the SCOZA cannot describe the formation of inhomogeneous microphases \cite{pini:06:0}.

Landau-Ginzburg-Wilson (LGW)\cite{zinn-justin:89:0,amit:84:0} theory turned out to be very successful in describing universal properties of  critical phenomena associated with transitions between uniform phases. Landau-Brazovskii (LB)\cite{brazovskii:75:0,fredrickson:87:0,podneks:96:0} theory on the other hand predicts universal properties of transitions between uniform and periodically ordered  lyotropic liquid crystals. 
The statistical field-theoretic methods are very powerful in determining effects of long-range correlations between fluctuations that lead to ordering - either to phase separation or to formation of various ordered structures on the nanometer length scale.  The effective Hamiltonians in Landau theories are given in terms of phenomenological parameters whose precise relation with thermodynamic and material properties  is irrelevant for the universal, i.e. substance independent properties. Unfortunately, the general, abstract theory has limited predicting power for particular systems.

On the other hand, in the very successful density functional theory \cite{evans:79:0} the short-range structure is taken into account quite precisely, whereas long- and intermediate range scale fluctuations are taken into account in a very crude mean-field approximation in most applications of the DFT. The DFT and LGW or LB theories are complementary in the treatment of the  short- and long range correlations.
It is thus desirable to develop DFT 
 with the form of the grand thermodynamic potential functional that includes the contribution associated with the mesoscopic scale fluctuations. In order to develop  such a theory one should 
perform a systematic  coarse-graining procedure with controllable accuracy, and derive effective Hamiltonians from the microscopic ones. Nonuniversal properties are correctly described within the collective variables (CV) \cite{yukhnovskii:58:0,yukhnovskii:78:0,caillol:06:0} and hierarchical reference (HRT) \cite{parola:85:0,parola:95:0} theories.  However, the HRT is restricted to critical phenomena, and the formal structure of these theories is rather complex.

Here we propose an alternative approach, resulting in a density functional theory with a rather simple structure.  The theory allows for including in the grand potential the contribution associated with long-range correlations between fluctuations. The latter contribution can be calculated within field-theoretic methods. Within the framework of our theory it is possible to determine phase transitions between different soft-crystalline and uniform phases. In special cases of phase separation or weak ordering into soft crystals the theory reduces to the standard LGW or LB theories respectively, with the coupling constants expressed in terms of density, temperature and the interaction potential. Within the present approach  phase diagrams in terms of density and temperature, rather than abstract phenomenological parameters can be obtained, and validity of the  LGW or LB theories in particular systems for given thermodynamic conditions can be verified. This kind of approach was applied already to highly charged colloids \cite{ciach:06:0} and to ionic systems \cite{ciach:00:0,ciach:01:0,ciach:01:1,ciach:03:0,ciach:03:1,ciach:06:1,ciach:06:2,patsahan:07:0,ciach:07:0}.

The derivation of the theory is described in sec.II. In sec.IIa the mesoscopic state (mesostate) is defined,  and in sec. IIb the probability distribution for the mesostates is derived from the microscopic theory. In sec.IIIa the correlation functions for the mesoscopic densities are introduced and their relation with the microscopic correlation functions is discussed. Vertex functions and the grand-potential density functional are introduced in sec. IIIb. Self-consistent equations for the two-point vertex  functions (related to direct correlation functions) are derived in sec. IIIc. Periodic structures are considered in sec. IIId, where the approximate expression for the grand potential is also given. In sec.IV the general framework of the theory derived in the preceding sections is applied to a particular approximation for the probability distribution for the mesostates, related to the local density approximation. The self-consistent equations for the two-point functions reduce in this approximation to simpler forms given in sec. IV a, and the approximation for the grand potential is given in sec. IVb. The relation between the present theory and the Landau-type theories is discussed in sec.IVd. Explicit results for the case of weak ordering are briefly described in sec. V. In sec. Va we study the $\lambda$-line in the Brazovskii theory. In sec. Vb we limit ourselves to universal features of the phase diagrams obtained in the simplest one-shell mean field (MF) approximation, and show the universal sequence of phases: disordered, bcc, hexagonal, lamellar, inverted hexagonal, inverted bcc and disordered for increasing density of particles (well below the close-packing density). The details of the phase diagrams can be obtained beyond the one-shell MF approximation. However, these details depend on the shape of the interaction potential.  Studies of particular systems go beyond the scope of this work and will be described elsewhere\cite{ciach:08:2}. For an illustration 
 we consider a particular form of so called short-range attraction long-range repulsion (SALR) potential that has drawn considerable interest recently \cite{sear:99:0,sciortino:04:0,sciortino:05:0,campbell:05:0,pini:06:0,candia:06:0,archer:07:0,archer:07:1,archer:08:0,archer:08:1}. In sec. Vc we quote results for a particular form of the SALR potential in the two-shell MF approximation, and show stability of the bicontinuous gyroid phase between the hexagonal and the lamellar phases. This phase may be related to the network-forming cluster of colloids observed experimentally \cite{campbell:05:0}.
\section{Coarse graining}
Let us consider hard spherical objects that  interact with arbitrary spherically-symmetric potentials. The approach can be generalized to hard object of different shapes, but it is easier to fix attention on spheres. A microstate is given by the sequence $\{{\bf r}_{\alpha}\}_{\alpha=1...N}$ describing the positions of the centers of $N$ spheres. In addition to the microscopic density,
\begin{equation}
\label{micror}
 \hat\rho({\bf r},\{{\bf r}_{\alpha}\}):=\sum_{\alpha}\delta({\bf r}-{\bf r}_{\alpha})
\end{equation}
 we consider the microscopic volume fraction 
\begin{equation}
\label{microe}
 \hat\eta({\bf r},\{{\bf r}_{\alpha}\}):=\sum_{\alpha}\theta(\sigma-|{\bf r}-{\bf r}_{\alpha}|)
\end{equation}
where
$\sigma$ is the diameter of the hard sphere. The energy in the microstate $ \{{\bf r}_{\alpha}\}_{\alpha=1...N}$  is given by
\begin{eqnarray}
 \label{microenergy}
E[\{{\bf r}_{\alpha}\}]=\sum_{\alpha>\beta}V(|{\bf r}_{\alpha}-{\bf r}_{\beta}|),
\end{eqnarray}
where $ V(|{\bf r}-{\bf r'}|)$ is the pair interaction potential. 
 \subsection{Mesoscopic density and mesoscopic volume fraction}
Let us  choose the mesoscopic length scale $R\ge \sigma/2$ and consider spheres $S_R({\bf r})$ of radius $R$ and centers at ${\bf r}$ that cover the whole volume $V$ of the system (for the bulk system we assume periodic boundary conditions). We define the {\it mesoscopic} density and the {\it mesoscopic} volume fraction at ${\bf r}$ by
\begin{equation}
\label{mesor}
 \rho({\bf r}):=\frac{1}{V_S} \int_{{\bf r'}\in S_R({\bf r})}\hat\rho({\bf r'},\{{\bf r}_{\alpha}\})
\end{equation}
and
\begin{equation}
\label{mesoe}
 \eta({\bf r}):=\frac{1}{V_S} \int_{{\bf r'}\in S_R({\bf r})}\hat\eta({\bf r'},\{{\bf r}_{\alpha}\})
\end{equation}
respectively,
where $V_S=4\pi R^3/3$. For brevity we shall use the notation $\int_{\bf r}\equiv \int d{\bf r}$, indicating the integration region $S$ by $\int_{{\bf r'}\in S}$ when necessary.

For given mesoscopic length scale $R$, and given microstate $\{{\bf r}_{\alpha}\}$, $\eta({\bf r})$ is a continuous field such that $\eta({\bf r})\le \eta_{cp}$ for all ${\bf r}$, where $\eta_{cp}$ is the close-packing volume fraction. Moreover, the gradient of $\eta$ is small, $|\nabla\eta|<1/R$ . The mesoscopic density has discontinuities. The steps are $n/V_S$, where  $n$ is the difference between the number of the centers of hard spheres included in $S_R({\bf r})$ and $S_R({\bf r}+d{\bf r})$.

\begin{figure}
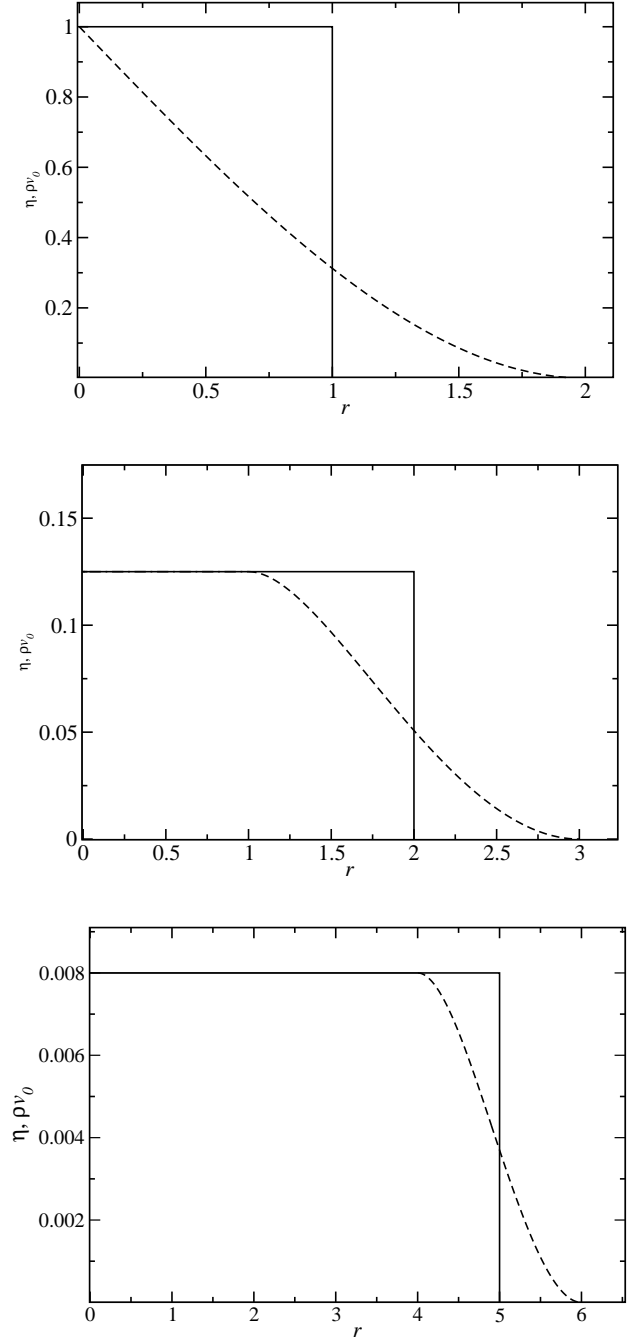

\includegraphics[scale=0.33]{fig1a.eps}\\
\vskip0.6cm
\includegraphics[scale=0.33]{fig1b.eps}\\
\vskip0.6cm
\includegraphics[scale=0.33]{fig1c.eps}
\caption{The mesoscopic volume-fraction $\eta$ defined in Eq.(\ref{mesoe}) (dashed lines) and $\rho v_0$  (solid lines), where $\rho$ is defined in Eq.(\ref{mesor}) and $v_0=\pi\sigma^3/6$,  for the microstate in which a single hard sphere of a radius $\sigma/2$ is located at ${\bf r}={\bf 0}$. Top panel: $2R/\sigma=1$. Central panel: $2R/\sigma=2$. Bottom panel: $2R/\sigma=5$. In this particularly simple case the fields (\ref{mesoe}) and (\ref{mesor}) are functions of the distance $r$ from the center of the hard sphere. In each case $\int_{\bf r}\eta({\bf r})=v_0$. The distance $r$ is in $\sigma/2$ units, $\eta$ and $\rho v_0$ are dimensionless.}
\end{figure}

The microstate $\{{\bf r}_{\alpha}\}$ for which (\ref{mesor}) (or (\ref{mesoe}))  holds for all ${\bf r}$ is called compatible with the field  $\rho({\bf r})$ (or $\eta({\bf r})$) for given $R$. In 
order to indicate that  the microstate $\{{\bf r}_{\alpha}\}$ is compatible with the field   $\rho({\bf r})$ (or $\eta({\bf r})$),
the notation $\{{\bf r}_{\alpha}\}\in \{\rho({\bf r}),R\}$ (or $\{{\bf r}_{\alpha}\}\in \{\eta({\bf r}),R\}$) will be used.
The set of all microstates compatible for given mesoscopic length scale $R$ with the field  $\rho({\bf r})$ (or $\eta({\bf r})$) is called a mesostate at the length scale $R$.

For given mesoscopic length scale $R$ the set of all microstates can be split into disjoint subsets, such that each subset contains all microstates compatible with a particular field  $\rho({\bf r})$ (or $\eta({\bf r})$) according to the  definition (\ref{mesor} (or (\ref{mesoe})), and no other microstates. 

The mesostate gives less detailed information about the state of the system than the microstate, but more detailed than the macrostate. 
 The mesostates depend on the chosen length scale (see Fig.1). For $R\to\infty$ the mesostates become identical with the macrostates. This is because so defined mesostate is characterized by the number density (or volume fraction of particles) in the whole system.
(Periodic boundary conditions are assumed). On the other hand, when $R=\sigma/2$, Eq.(\ref{mesoe}) defines the one-to-one relation between the microstates and the mesostates. 
To prove the above let us 
 consider two different microstates $\{{\bf r}_{\alpha}\}$ and $\{{\bf r'}_{\beta}\}$. If they are different, then a center of at least one sphere in the first microstate does not coincide with a center of any sphere in the second microstate. Let us assume that ${\bf r}_1\ne {\bf r'}_{\beta}$ for all $\beta =1,...,N$. Let us calculate (\ref{mesoe}) for ${\bf r}={\bf r}_1$. For the first microstate we get $\eta({\bf r})=1$, for the second microstate $\eta({\bf r})\ne 1$. Thus, different microstates cannot be compatible with the same field defined in (\ref{mesoe}).

The choice of $R$ depends on the problem under consideration, in particular on the length scale $\lambda$ characteristic for ordering, and is determined by the Hamiltonian. In order to describe the ordering on the length scale $\lambda$, one should choose $\sigma/2<R<\lambda/2$. 
\subsection{Probability distribution for the mesostates}
Let us calculate the probability density that the mesostate  $\{\rho({\bf r}),R\}$ (or $\{\eta({\bf r}),R\}$) occurs in the system spontaneously. This is equal to the probability that any microstate compatible with   $\rho({\bf r})$ (or $\eta({\bf r})$) occurs. We derive the expressions in terms of $\rho$; in terms of  $\eta$ the  theory has the same formal structure. In an open system in contact with the thermostat  the probability density of the mesostate $ \{\rho,R\}$ is given by
\begin{equation}
\label{p}
 p[\rho]=\frac{1}{\Xi}\int_{\{{\bf r}_{\alpha}\}\in\{\rho,R\}}e^{-\beta(H-\mu\int_{\bf r}\rho({\bf r}))}
\end{equation}
where $H$ is the microscopic Hamiltonian, $\beta=1/(k_BT)$  and  $ \int_{\{{\bf r}_{\alpha}\}\in\{\rho,R\}}$ is the symbolic notation for the integration over all microstates compatible with $\rho$.  $\mu$ and $T$ are the chemical potential and temperature respectively, and $\beta=1/k_BT$, with $k_B$ denoting the Boltzmann constant. Finally,
\begin{equation}
\label{Xi1}
 \Xi=\int_{\{{\bf r}_{\alpha}\}}e^{-\beta(H-\mu\int_{\bf r}\rho({\bf r}))}=
\int^{'} D\rho \int_{\{{\bf r}_{\alpha}\}\in\{\rho,R\}}e^{-\beta(H-\mu\int_{\bf r}\rho({\bf r}))}.
\end{equation}
The functional integral $ \int^{'} D\rho $ in (\ref{Xi1}) is over all mesostates $ \{\rho,R\}$, which is indicated by the prime.

Fixing the mesostate in the system is equivalent to the constraint on the microstates of the form (\ref{mesor}). In the presence of the constraint $ \{\rho,R\}$ the grand potential is denoted by $\Omega_{co}[\rho]$ and is given by
\begin{equation}
\label{Omcod}
 e^{-\beta\Omega_{co}[\rho]}=\int_{\{{\bf r}_{\alpha}\}\in\{\rho,R\}}e^{-\beta(H-\mu\int_{\bf r}\rho({\bf r}))}.
\end{equation}
Thus, the probability density of a spontaneous occurrence of the mesostate  $ \{\rho,R\}$ (equal to the probability density that any of the microstates compatible with $\rho$ occurs) is given by
 \begin{equation}
\label{p2}
  p[\rho]=\frac{e^{-\beta\Omega_{co}[\rho]}}{\Xi} .
 \end{equation}
where
\begin{equation}
\label{Xi2}
 \Xi=\int^{'} D\rho  e^{-\beta\Omega_{co}[\rho]}.
\end{equation}

We obtain a mesoscopic theory with the same structure as the standard statistical mechanics. The integration over all microstates is replaced in (\ref{Xi2}) by the integration over all mesostates. The Hamiltonian is replaced in (\ref{p2}) and (\ref{Xi2}) by the grand potential in the presence of the constraint of compatibility with the given mesostate that is imposed on the microstates. The above formulas are exact. So far we just rearranged the summation over microstates. In (\ref{Xi2}) the integration over microstates compatible with each mesostate is included in $\Omega_{co}$, and then we perform a summation over all mesostates. The reason for doing so is the possibility of treating the summation over the mesostates and over the microstates compatible with a particular mesostate on different levels of approximation.
In a similar way one can define a mesoscopic theory for the mesoscopic volume fraction. Both approaches are equivalent. The advantage of the density field is the fact that the microscopic and functional-density theories are based on the density rather than on the volume fraction, and the form of $\Omega_{co}$ should be determined within microscopic theories. The advantage of the volume fraction is the fact that it is a continuous field. 

Note that in the mesoscopic theory the mesostate $\{\rho_0,R\}$ that corresponds to the global minimum of $ \Omega_{co}[\rho]$  is analogous to the ground state in the microscopic theory (similar role plays $\eta_0$ 
corresponding to the minimum of $\Omega_{co}[\eta]$). This is because the ground state is the  microstate that corresponds to the global minimum of the Hamiltonian. In this context the important property of the ground state is the fact that it corresponds to \textit{the most probable single microstate}. Likewise $\{\rho_0,R\}$ is \textit{the single mesostate that occurs with the highest probability}, because (\ref{p2}) assumes a maximum when $ \Omega_{co}[\rho]$ assumes a minimum. 
\section{Grand potential and the correlation functions for the mesoscopic densities}
The grand potential in the system subject to the constraint for the mesoscopic density distribution $\rho({\bf r})$ can be written in the form 
 \begin{equation}
\label{Omco}
\Omega_{co}=U-TS-\mu N,
\end{equation}
 where $U,S, N$ are the internal energy, entropy and the number of molecules respectively in the system with the constraint  (\ref{mesor}) or (\ref{mesoe}) imposed on the microscopic densities or the volume fractions respectively.  $U$ is given by the well known expression
\begin{equation}
\label{U}
 U[\rho^*]=\frac{1}{2}\int_{\bf r_1}\int_{\bf r_2}V_{co}({\bf r}_1-{\bf r}_2)\rho^*({\bf r}_1)\rho^*({\bf r}_2),
\end{equation}
where 
\begin{eqnarray}
\label{Vco}
 V_{co}({\bf r}_1-{\bf r}_2)= V(r_{12})g_{co}({\bf r}_1-{\bf r}_2),
\end{eqnarray}
 $r_{12}=|{\bf r}_1-{\bf r}_2|$, and $g_{co}({\bf r}_1-{\bf r}_2)$   is the microscopic pair correlation function in the system with the constraint (\ref{mesor})  imposed on the microscopic states. From now on we  consider dimensionless density 
\begin{equation}
 \label{dimless}
\rho^*=\rho\sigma^3.
\end{equation}

In the case of the considered systems (no internal degrees of freedom of the particles) the entropy $S$ satisfies the relation $-TS=F_h$, where $F_h$ is the free-energy of the reference hard-sphere system with the constraint  (\ref{mesor}) or (\ref{mesoe}) imposed on the microscopic  densities   or the microscopic volume fractions respectively. 
\subsection{Correlation functions for the mesoscopic densities and their generating functional}
 Let us introduce an external field $J({\bf r})$ and the grand-thermodynamic potential functional 
\begin{equation}
\label{OmJ}
 \Omega[\beta J]:=-k_BT\log\Big[\int^{'}D\rho^*  e^{-\beta[\Omega_{co}[\rho^*]-
\int_{\bf r}J({\bf r})\rho^*({\bf r})]}\Big].
\end{equation}

The generating functional for the (connected) correlation functions for the mesoscopic densities is $-\beta\Omega[\beta J]$, and
\begin{equation}
\label{genfu}
 \langle\rho^*({\bf r}_1)...\rho^*({\bf r}_n)\rangle^{con}=
\frac{\delta^n(-\beta\Omega[\beta J])}{\delta(\beta J({\bf r}_1))...\delta(\beta J({\bf r}_n))}.
\end{equation}
In particular, the above gives
\begin{equation}
\label{<r>}
 \langle\rho^*({\bf r})\rangle=\frac{\int^{'}D\rho^*  e^{-\beta[\Omega_{co}[\rho^*]-
\int_{\bf r}J({\bf r})\rho^*({\bf r})]}\rho^*({\bf r})}{\int^{'}D\rho  e^{-\beta[\Omega_{co}[\rho^*]-
\int_{\bf r}J({\bf r})\rho^*({\bf r})]}}=\frac{\sigma^3}{V_S}\int_{{\bf r'}\in S({\bf r})}\langle \hat\rho({\bf r}')\rangle,
\end{equation}
where (\ref{mesor}), (\ref{Omcod}) and (\ref{Xi1}) were used. We use the same notation $\langle ...\rangle$ for the microscopic and the mesoscopic average.
The average value of the \textit{mesoscopic} density at the point ${\bf r}$ is the average \textit{microscopic} density \textit{integrated} over the sphere $S_R({\bf r})$ of the radius $R$, and divided by its volume $V_S$. Clearly, $\langle\rho^*({\bf r})\rangle$ depends on $R$. If, however, $\langle\rho^*({\bf r})\rangle$  is independent of $R$ for some range of $\sigma/2<R<\lambda/2$,  then this function gives information about actual ordering on the  length scale  $ \lambda$, because the average value of the mesoscopic density  is the microscopic density averaged over regions smaller than the length scale on which the ordering occurs. 
 
For the correlation function for the mesoscopic density  we introduce the notation
\begin{eqnarray}
\label{corfuecon}
{\cal G}_2({\bf r}_1-{\bf r}_2)= \langle\rho^*({\bf r}_1)\rho^*({\bf r}_2)\rangle^{con},
\end{eqnarray}
where here and below the superscript $con$ denotes
\begin{eqnarray}
 \langle A(\rho^*({\bf r}_1))B(\rho^*({\bf r}_2))\rangle^{con}=\langle A(\rho^*({\bf r}_1))B(\rho^*({\bf r}_2))\rangle-\langle A(\rho^*({\bf r}_1))\rangle\langle B(\rho^*({\bf r}_2))\rangle,
\end{eqnarray}
and where
\begin{eqnarray}
\label{corfue}
 \langle\rho^*({\bf r}_1)\rho^*({\bf r}_2)\rangle=
\frac{\int^{'}D\rho  e^{-\beta[\Omega_{co}[\rho^*]-
\int_{\bf r}J({\bf r})\rho^*({\bf r})]}\rho^*({\bf r}_1)\rho^*({\bf r}_2)}{\int^{'}D\rho  e^{-\beta[\Omega_{co}[\rho^*]-
\int_{\bf r}J({\bf r})\rho^*({\bf r})]}}=
\\
\nonumber
\frac{\sigma^3}{V_S}\int_{{\bf r'}\in S({\bf r}_1)}\frac{\sigma^3}{V_S}\int_{{\bf r''}\in S({\bf r}_2)}\langle \hat\rho({\bf r}')\hat\rho({\bf r}'')\rangle.
\end{eqnarray}

The correlation function for the {\it mesoscopic} density  at ${\bf r}_1$ and ${\bf r}_2$ is the correlation function for the {\it microscopic}  density  at the points ${\bf r}'\in S_R({\bf r}_1)$ and ${\bf r}''\in S_R({\bf r}_2)$, \textit{integrated over the spheres} $S_R({\bf r}_1)$ and $S_R({\bf r}_2)$ respectively.   While the \textit{ microscopic} pair distribution function $g_2({\bf r}'-{\bf r}'') =g({\bf r}'-{\bf r}'')\rho({\bf r}')\rho({\bf r}'')=\langle \hat\rho({\bf r}')\hat\rho({\bf r}'')\rangle-\langle \hat\rho({\bf r}')\rangle\delta({\bf r}'-{\bf r}'')$ vanishes for $|{\bf r}'-{\bf r}''|<\sigma$, the \textit{mesoscopic} quantity defined by Eqs.(\ref{corfuecon})-(\ref{corfue}) does not vanish for  ${\bf r}_1={\bf r}_2$ when $R>\sigma/2$ (see Fig.2).
 \begin{figure}
\includegraphics[scale=0.35]{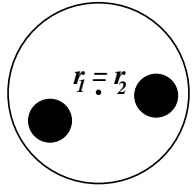}
\caption{Microscopic configuration contributing to the correlation function ${\cal G}_2({\bf r}_1-{\bf r}_2)$ for ${\bf r}_1={\bf r}_2$ in the case of $R>\sigma/2$. The black circles represent particles, and the open circle represents the coinciding spheres $S_R({\bf r}_1)=S_R({\bf r}_2)$, over which the microscopic correlations are averaged. Note that the black spheres do not overlap, and their centers are both included in $S_R$. Thus, the microscopic correlation function for this microstate does not vanish.}
\end{figure}
 The larger $R$ is, the larger  is $ \langle\rho({\bf r}_1)\rho({\bf r}_2)\rangle-\langle \rho({\bf r}_1)\rangle\delta({\bf r}_1-{\bf r}_2)$ for ${\bf r}_1\to {\bf r}_2$. 

\subsection{Grand-potential functional of the mesoscopic density, and the vertex functions}
Let us introduce the Legendre transform
\begin{equation}
 \label{lege}
\beta F[\bar\rho^*]:=\beta\Omega[\beta J]+\int_{\bf r}\beta J({\bf r})\bar\rho^*({\bf r})
\end{equation}
where 
\begin{equation}
\label{Phi}
 \bar\rho^*({\bf r})=\frac{\delta(-\beta\Omega)}{\delta(\beta J({\bf r}))}
\end{equation}
is the dimensionless average field  for given $J({\bf r})$. The equation of state  takes the form
\begin{equation}
 \label{EOS}
\frac{\delta (\beta F)}{\delta\bar\rho^*({\bf r})}=\beta J({\bf r}).
\end{equation}

In general $\bar\rho^*$ may differ from any mesostate defined by (\ref{mesor}).
Let  the functional $\Omega_{co}$, defined  for the mesostates, be extended beyond the set of the mesostates. Let the extension be
 defined in (\ref{Omco}) on the Hilbert space of fields $f$ that fulfill the restrictions following from the properties of the mesostates, and let us keep the notation $\Omega_{co}$ for this extension.  The key restriction on the mesoscopic volume fraction is the gradient, whose upper limit is 
$ \sigma^2/(2R)^3 <  1/(2R)$. The mesoscopic number density has discontinuities, with the steps $\sim \sigma^3/R^3$ decreasing for increasing $R/\sigma$ (see Fig.1), and the separation between the steps is $\sim \sigma$. In the approximate theory we shall consider the Hilbert space of  fields 
such that in Fourier representation $\tilde f({\bf k })=0$ for the wavenumbers  $k\ge \pi/R $. 
Here and below
$\tilde f({\bf k})$ is the Fourier transform of $f({\bf r})$.
 The mesoscopic volume fraction and density are bounded from above, and  close-packing is the natural limit for their values.  However, the fields $\tilde\rho^*({\bf k})$ with large values yield large values of $\Omega_{co}$ (see (\ref{Omco}) and  (\ref{U})), and in turn small values of the Boltzmann factor (\ref{p2}), therefore inclusion of such fields in  the Hilbert space should not have large effect on the results. 
Selfconsistency of the approach requires that the average density obtained within this theory is bounded from above by the close-packing density.

We can define
\begin{eqnarray}
\label{Hfl}
H_{fluc}[\bar\rho^*,\phi]=\Omega_{co}[\bar\rho^*+\phi]-\Omega_{co}[\bar\rho^*]=
\\
\nonumber
\sum_{n=1}\int_{\bf r_1}...\int_{\bf r_n}\frac{{\cal C}^{co}_n({\bf r}_1,...{\bf r}_n|\bar\rho^*]}{n!}\phi({\bf r}_1)...\phi({\bf r}_n)
\end{eqnarray}
where $\bar\rho^*$ is given in (\ref{Phi}) and 
$ {\cal C}^{co}_n({\bf r}_1,...{\bf r}_n|\bar\rho^*]$ is the $n-$th functional derivative of $ \Omega_{co}[\rho^*]$ at $\rho^*=\bar\rho^*$. The second equality holds for the extended functional, for which the derivative can be defined. By definition $\langle \phi\rangle=0$. 
Then from (\ref{OmJ}) and (\ref{lege})  we obtain
\begin{equation}
\label{Omm}
 -\beta\Omega[\beta J]=-\beta\Omega_{co}[\bar\rho^*]+\int_{\bf r}\beta J({\bf r})\bar\rho^*({\bf r})+
\log\Big[\int D\phi e^{-\beta[H_{fluc}-\int_{\bf r} J({\bf r})\phi({\bf r})]}\Big]
\end{equation}
and 
\begin{equation}
\label{FF}
 \beta F[\bar\rho^*]=\beta\Omega_{co}[\bar\rho^*]-\log\Big[\int D\phi e^{-\beta[H_{fluc}-\int_{\bf r} J({\bf r})\phi({\bf r})]}\Big].
\end{equation}

Eq. (\ref{FF}) defines the functional in which the microscopic scale fluctuations, with frozen fluctuations of the mesoscopic density over the length scales larger than $R$, are included in the first term, and the mesoscopic scale fluctuations are included in the last term. The expression (\ref{FF}) would be exact, if the functional integration were restricted to the mesostates. However, the powerful methods of the functional analysis could not be applied. There are no easy ways of estimating the error associated with inclusion in the fluctuation contribution fields that do not represent the actual mesostates.
The  Eq.(\ref{FF})  should serve as a starting point for various approximate theories. The idea is to apply approximate microscopic theories to the first term, and field-theoretic approaches to the second term.
\subsection{Equations for the correlation functions}
Note that from (\ref{FF}) it follows that the vertex  functions (related to the direct correlation functions) defined by
 \begin{equation}
\label{calCn}
 {\cal C}_n({\bf r}_1,...,{\bf r}_n)= \frac{\delta^n \beta F[\bar\rho^*]}{\delta\bar\rho^*({\bf r}_1)...\delta\bar\rho^*({\bf r}_n)}
 \end{equation}
consist of two terms: the first one is the contribution from the fluctuations on the microscopic length scale ($<R$) with frozen fluctuations on the mesoscopic length scale, the second one is the contribution from the fluctuations on the mesoscopic length scale ($>R$).

The average density for $J=0$ satisfies the equation (see Eqs.(\ref{EOS}) and (\ref{FF}))
\begin{eqnarray}
\label{avdg}
 \frac{\delta\beta\Omega_{co}[\bar\rho^*]}{\delta\bar\rho^*({\bf r}) }+\langle \frac{\delta(\beta H_{fluc})}{\delta\bar\rho^*({\bf r})}\rangle=0,
\end{eqnarray}
where the averaging is over the fields $\phi$ with the probability $\propto\exp(-\beta H_{fluc}[\bar\rho^*,\phi])$.
For the two-point function from (\ref{calCn}) and (\ref{FF}) we obtain the equation
\begin{eqnarray}
\label{CalC2}
 2{\cal C}_2({\bf r}_1,{\bf r}_2)= {\cal C}_2^{co}({\bf r}_1,{\bf r}_2)+\langle \frac{\delta^2(\beta H_{fluc})}{\delta\bar\rho^*({\bf r}_1)\delta\bar\rho^*({\bf r}_2)}\rangle -\langle \frac{\delta(\beta H_{fluc})}{\delta\bar\rho^*({\bf r}_1)}\frac{\delta(\beta H_{fluc})}{\delta\bar\rho^*({\bf r}_2)}\rangle^{con}+\\
\nonumber
\int_{\bf r'}\Bigg[\langle \frac{\delta H_{fluc}}{\delta\bar\rho^*({\bf r}_1)}\phi({\bf r}')\rangle {\cal C}_2({\bf r}',{\bf r}_2)+\langle \frac{\delta H_{fluc}}{\delta\bar\rho^*({\bf r}_2)}\phi({\bf r}')\rangle {\cal C}_2({\bf r}',{\bf r}_1)
\Bigg].
\end{eqnarray}
In calculating the functional derivative of $F$ (Eq.(\ref{FF})) we used the equality  $\delta J({\bf r}_1)/\delta\bar\rho^*({\bf r}_2) ={\cal C}_2 ({\bf r}_1,{\bf r}_2)$ (see (\ref{EOS})).   From (\ref{Hfl}) we have
\begin{eqnarray}
\label{der}
 \frac{\delta^m(\beta H_{fluc})}{\delta\bar\rho^*({\bf r})...\delta\bar\rho^*({\bf r}^{(m)})}=\sum_{n=1}^{\infty}\int_{\bf r_1}...\int_{\bf r_n}\frac{{\cal C}^{co}_{n+m}({\bf r}_1,...{\bf r}_n,{\bf r},...,{\bf r}^{(m)}|\bar\rho^*]}{n!}\phi({\bf r}_1)...\phi({\bf r}_n).
\end{eqnarray}
Higher order vertex functions can be obtained from Eqs.(\ref{calCn}) and (\ref{FF}) in a similar way. As a result, a hierarchy of equations where the vertex functions are expressed in terms of the many-body correlation functions is obtained.

Eqs. (\ref{avdg})-(\ref{der}) allow for including the fluctuation contribution to the  vertex  functions on different levels of approximation. In the mean-field approximation the fluctuation contribution is just neglected. In the lowest-order nontrivial approximation  only terms proportional to the pair-correlation function $\langle\phi({\bf r})\phi({\bf r}')\rangle$ are included in (\ref{avdg})-(\ref{CalC2}),
by truncating the expansion in (\ref{der}) at $n=2$. Then Eqs.(\ref{avdg}) and (\ref{CalC2}) assume the approximate forms
\begin{eqnarray}
\label{avdgg}
 \frac{\delta\beta\Omega_{co}[\bar\rho^*]}{\delta\bar\rho^*({\bf r}) }+\int_{\bf r_1}\int_{\bf r_2}{\cal G}_2({\bf r}_1,{\bf r}_2)
{\cal C}_3^{co}({\bf r}_1,{\bf r}_2,{\bf r})=0
\end{eqnarray}
and
\begin{eqnarray}
 \label{CalC2g}
2{\cal C}_2({\bf r}_1,{\bf r}_2)= 3{\cal C}_2^{co}({\bf r}_1,{\bf r}_2)+\int_{\bf r}\int_{\bf r'}
{\cal G}_2({\bf r},{\bf r}')\Big[ \frac{{\cal C}_4^{co}({\bf r},{\bf r}',{\bf r}_1,{\bf r}_2)}{2}-{\cal C}^{co}_2({\bf r},{\bf r}_1) {\cal C}_2^{co}({\bf r}',{\bf r}_2)\Big].
\end{eqnarray}

From (\ref{EOS}) and (\ref{genfu}) we obtain the analog of the Ornstein-Zernicke equation
\begin{equation}
\label{OZ}
 {\cal C}_2({\bf r}_1,{\bf r}_2)= \frac{\delta^2 (\beta F)}{\delta\bar\rho^*({\bf r}_2)\delta\bar\rho^*({\bf r}_1)}=\Bigg[
\frac{\delta \bar\rho^*({\bf r}_2)}{\delta\beta J({\bf r}_1)} \Bigg]^{-1}={\cal G}_2({\bf r}_1,{\bf r}_2)^{-1},
\end{equation}
or in the equivalent form
\begin{equation}
 \label{OZ1}
\int_{\bf r_2}{\cal C}_2({\bf r}_1,{\bf r}_2){\cal G}_2({\bf r}_2,{\bf r}_3)=\delta({\bf r}_1-{\bf r}_3).
\end{equation}
We assume that the form of $\Omega_{co}$ is known from the microscopic theory, and consider it as an input to the mesoscopic theory.
Selfconsistent solutions of  Eqs.(\ref{CalC2g})-(\ref{OZ1}) yield  the two-point vertex and  correlation functions, ${\cal C}_2$ and $ {\cal G}_2$ respectively,  for given forms of ${\cal C}_n^{co}$.
Note that Eq.(\ref{CalC2g}) plays a role of the closure to the OZ equation (\ref{OZ1}).

Eq. (\ref{avdgg}) is the minimum condition for the grand potential. Since there may exist several local minima, the solution corresponds to a stable or to a metastable phase when the grand potential assumes the global or the local minimum. The solution of Eq.(\ref{avdgg}) corresponding to the global minimum gives the average density for given $\mu$ and $T$ in the lowest nontrivial order beyond MF.
\subsection{Periodic structures; general case} 
Let us consider  periodic density profiles
\begin{eqnarray}
\label{perdens}
 \bar\rho^*({\bf r})=\bar\rho^*_0+\Phi({\bf r})
\end{eqnarray}
where
\begin{eqnarray}
 \Phi({\bf r}+{\bf P})=\Phi({\bf r})
\end{eqnarray}
and ${\bf P}=\sum_{i}^3n_i{\bf p}_i$ where ${\bf p}_i$ are the vectors connecting the centers of the nearest-neighbor unit cells and  $n_i$ are integer. The $\bar\rho^*_0$ is the space-averaged density, i.e.
\begin{eqnarray}
\label{Pp}
 \int_{{\bf r}\in {{\cal V}_u}}\Phi({\bf r})=0,
\end{eqnarray}
where ${\cal V}_u$ is the unit cell of the periodic structure, whose volume is denoted by $V_u$.
In the case of  periodic structures 
\begin{eqnarray}
\label{CP}
{\cal C}_2({\bf r}_1+{\bf P},{\bf r}_2+{\bf P})={\cal C}_2({\bf r}_1,{\bf r}_2)={\cal C}_2(\Delta{\bf r}|{\bf r}_2)
\end{eqnarray}
 where 
$\Delta{\bf r}={\bf r}_1-{\bf r}_2\in R^3$ and ${\bf r}_2\in {\cal V}_u$.  

Let us consider the Gaussian functional 
\begin{eqnarray}
\label{Hagac}
 {\cal H}_G[\bar\rho^*,\phi]= \frac{1}{2}\int_{{\bf r}_1}\int_{{\bf r}_2}\phi({\bf r}_1) {\cal C}_2({\bf r}_1-{\bf r}_2|{\bf r}_2 )\phi({\bf r}_2)=
\frac{1}{2}\int_{\bf k}\int_{\bf k'}\tilde \phi({\bf k})\tilde {\cal C}_2({\bf k},{\bf k}+{\bf k}')\tilde\phi({\bf k}').
\end{eqnarray}
Here and below we use the simplified notation
 $\int_{\bf k}=\int d{\bf k}/(2\pi)^3$. 
For periodic structures we have (see Appendix A)
\begin{eqnarray}
 \tilde {\cal C}_2({\bf k},{\bf k}+{\bf k}')=
\delta({\bf k}+{\bf k}')\tilde C_2({\bf k})
\end{eqnarray}
where 
\begin{eqnarray}
\label{CCga}
 \tilde C_2({\bf k})=\int_{\Delta{\bf r}} C_2(\Delta{\bf r})e^{i{\bf k}\cdot\Delta{\bf r}},
\end{eqnarray}
and 
\begin{eqnarray}
\label{CCga1}
 C_2(\Delta{\bf r})=\frac{1}{V_u}\int_{{\bf r_2}\in {\cal V}_u}{\cal C}_2(\Delta{\bf r}|{\bf r}_2)
\end{eqnarray}
is the vertex function averaged over the unit cell.
For periodic structures the functional (\ref{Hagac}) can be rewritten in the equivalent form
\begin{eqnarray}
\label{Haga}
 {\cal H}_G[\bar\rho^*,\phi]=
\frac{1}{2}\int_{\bf k}\tilde \phi({\bf k})\tilde C_2({\bf k})\tilde\phi(-{\bf k}).
\end{eqnarray}

From the analog of the Ornstein-Zernicke equation (\ref{OZ}) we obtain the relation
\begin{equation}
\label{OZF}
 \tilde C_2({\bf k})\tilde G_2({\bf k})=1
\end{equation}
between the vertex function averaged over the unit cell, and  
\begin{eqnarray}
\label{GGga}
 \tilde G_2({\bf k})=\int_{\bf r_1-r_2}\frac{1}{V_u}\int_{{\bf r}_1\in{\cal V}_u}e^{i{\bf k}\cdot({\bf r_1-r_2})}{\cal G}_2({\bf r}_1,{\bf r}_2)=\int_{\bf r_1-r_2}e^{i{\bf k}\cdot({\bf r_1-r_2})}G_2({\bf r_1-r_2}).
\end{eqnarray}
  The function $G_2({\bf r_1-r_2})$ describes the correlations between the density fluctuations at the distance ${\bf r_1-r_2}$, with the first point position averaged over the unit cell. 

In order to calculate the second term in (\ref{FF}) we  follow the standard field-theoretic procedure and write
\begin{equation}
\label{Har}
H_{fluc}[\bar\rho^*,\phi]
={\cal H}_{G}[\bar\rho^*,\phi]+\Delta{\cal H}[\bar\rho^*,\phi]
\end{equation}
where
${\cal H}_{G}[\bar\rho^*,\phi]$ is the Gaussian functional (\ref{Haga}).
Next we make an assumption that $\Delta{\cal
 H}[\bar\rho^*,\phi]$ can be treated as a small perturbation. When such an
 assumption is valid, we obtain \cite{podneks:96:0,ciach:06:1}
\begin{eqnarray}
\label{OmHarval}
\beta\Omega[\bar\rho^*]\approx \beta\Omega_{co}[\bar\rho^*]-\log\int D\phi 
e^{-\beta{\cal H}_G}
+\langle \beta\Delta{\cal H}\rangle_G
+O(\langle \beta\Delta{\cal H}\rangle_G^2).
\end{eqnarray}
where $\langle ...\rangle_G$ denotes averaging with the Gaussian Boltzmann
factor $e^{-\beta{\cal H}_G}$. 
Note that the Gaussian functional integrals can be calculated analytically. Eqs. (\ref{CalC2g}) - (\ref{OmHarval}) allow for calculation of the fluctuation contribution to the grand potential in the lowest nontrivial order when the form of $\Omega_{co}$ is known, and  the form of $\tilde C_2$ ((\ref{CCga}) and (\ref{CCga1})) is determined by a self-consistent solution of Eqs.(\ref{CalC2g}) and (\ref{OZ}). In general, each contribution to Eq.(\ref{OmHarval}) depends on the mesoscopic length scale $R$, but the $R$-dependent contributions must cancel against each other to yield  $R$-independent $\Omega$. 
By minimizing the density functional (\ref{OmHarval}) we find the equilibrium structure. 
\section{Approximate theory: local density approximation for $\Omega_{co}$}
In our theory different levels of approximation for $\Omega_{co}$  are possible. 
For an illustration  we demonstrate  the mesoscopic theory with the microscopic degrees of freedom  considered in a very crude approximation. In this approximation calculations are greatly simplified and analytical methods can be used. Next we show how the functional can be further reduced to the Landau-type form, and discuss in what cases such a reduction is justified. The expressions for the phenomenological parameters in the Landau-type theory are given in terms of the thermodynamic variables and $V_{co}$. 
\subsection{Derivation of the functional $\Omega_{co}$ in the local density approximation}
In order to derive an approximate form of $\Omega_{co}$ (Eq.(\ref{Omco})) we need an approximate form of the microscopic pair correlation function.
The microscopic correlation function has the limiting behavior
\begin{eqnarray}
 g_{co}({\bf r}_1-{\bf r}_2)=0 \hskip1cm {\rm for} \hskip1cm  |{\bf r}_1-{\bf r}_2|<\sigma \\
\nonumber
g_{co}({\bf r}_1-{\bf r}_2)\to 1 \hskip1cm {\rm for} \hskip1cm |{\bf r}_1-{\bf r}_2|\to \infty.
\end{eqnarray}
If the ordering in the system occurs on the length scale larger than $R\gg\sigma/2$, then the precise form of $g_{co}$ is not crucial and in the simplest approximation we assume
\begin{eqnarray}
\label{gtheta}
  g_{co}({\bf r}_1-{\bf r}_2)=\theta(|{\bf r}_1-{\bf r}_2|-\sigma).
\end{eqnarray}
 
 In order to develop an approximation for the entropy $S$ in the presence of the constraint imposed on the density profile,
let us  consider the number of the microstates  associated with different positions of the centers of particles included in the sphere $S_R({\bf r})$ for given $\rho({\bf r})$. We  assume that the corresponding contribution to the entropy  only weakly depends on the mesoscopic density at ${\bf r'}$ when $ |{\bf r}-{\bf r'}|\gg R$. When this assumption is satisfied, then the local density approximation can be applied, and
\begin{eqnarray}
\label{Fhfh}
 F_h[\rho^*]=\int_{\bf r}f_h(\rho^*({\bf r})),
\end{eqnarray}
where $f_h(\rho^*)$ is the free-energy density of the hard-sphere system of density $\rho^*=\rho \sigma^3$. 
For the latter we may assume the Percus-Yevick or the Carnahan-Starling approximation. We choose the former case (compressibility route) and assume
\begin{eqnarray}
\label{PY}
\beta f_h(\rho^*)=\rho^*\ln(\rho^*)-\rho^*+
\rho^*\Bigg[\frac{3\eta(2-\eta)}{2(1-\eta)^2}-\ln(1-\eta)\Bigg].
\end{eqnarray}
In this simple approximation we obtain the $R$-independent functional
\begin{eqnarray}
\label{Omcoap}
 \beta\Omega_{co}[\rho^*]=\frac{1}{2}\int_{{\bf r}_1}\int_{{\bf r}_2}\beta V_{co}(r_{12})
\rho^*({\bf r}_1)\rho^*({\bf r}_2)-\int_{\bf r}\beta f_h(\rho^*({\bf r}))-\int_{\bf r}\beta\mu\rho^*({\bf r}),
\end{eqnarray}
where $ V_{co}(r_{12})$ is defined in Eq.(\ref{Vco}),
and for $g_{co}(r_{12})$ we make the assumption (\ref{gtheta}).
Note that in this approximation the dominant contribution to the fluctuation part in (\ref{OmHarval}) should be independent of $R$, otherwise  the theory is not valid. This is because $\Omega$ cannot depend on the arbitrary length scale $R$. 

For  $F_h[\rho^*]$ given by (\ref{Fhfh}), i.e. for the local density approximation, the functional derivatives of the functional $\Omega_{co}$ take the forms
\begin{eqnarray}
 {\cal C}^{co}_n({\bf r}_1,...{\bf r}_n|\bar\rho^*]= \left\{ 
\begin{array}{lll}
\delta({\bf r}_1-{\bf r}_2)...\delta({\bf r}_{n-1}-{\bf r}_n)
\beta f_h^{(n)}(\bar\rho^*({\bf r}_1)) &\;\; {\rm for} &\;\; n\ge 3\\ 
\beta( f_h^{(2)}(\bar\rho^*({\bf r}_1))\delta({\bf r}_1-{\bf r}_2) +V_{co}(r_{12}))  &\;\; {\rm for}&\;\; n=2\\
\beta\Big(f_h^{(1)}(\bar\rho^*({\bf r}_1))-\mu +\int_{{\bf r}_2}V_{co}(r_{12})\bar\rho^*({\bf r}_2)\Big)&\;\; {\rm for}&\;\; n=1.
\end{array}
\right.
 \label{CoPY}
\end{eqnarray}
where $f_h^{(n)}(\bar\rho^*({\bf r}))$ denotes the $n$-th derivative of $f_h$ with respect to its argument, calculated at $\rho^*({\bf r})=\bar\rho^*({\bf r})$. In the local density approximation Eq.(\ref{Hfl}) assumes the simpler form, 
\begin{eqnarray}
\label{DelOM}
 \beta H_{fluc}[\bar\rho^*,\phi]=\beta\Omega_{co}[\bar\rho^*+\phi]-\beta\Omega_{co}[\bar\rho^*]=
\\
\nonumber
\frac{1}{2}\int_{\bf r_1}\int_{\bf r_2}\phi({\bf r}_1) {\cal C}^{co}_2({\bf r}_1,{\bf r}_2)\phi({\bf r}_2)
+\int_{\bf r} {\cal C}^{co}_1({\bf r})\phi({\bf r})+
\sum_{n=3}\int_{\bf r}\frac{\beta f_h^{(n)}(\bar\rho^*({\bf r}))}{n!}\phi({\bf r})^n .
\end{eqnarray}

The dominant contribution to the second term in (\ref{FF}) comes from small fields $\phi$; moreover, fields with large values do not represent the actual mesostates. For fields with small values the expansion in (\ref{DelOM}) can be truncated. For stability reasons the $\phi^4$ term must be included. When the expansion in (\ref{DelOM}) is truncated at the fourth order term, we obtain $ H_{fluc}$ of the form similar to the effective Hamiltonian in the Landau-type $\varphi^4$ theories, except that in nonuniform phases the coefficients that multiply $\phi^n$ depend on space position in a way determined by the form of $f_h(\bar\rho^*({\bf r}))$.
\subsection{Approximate equations for the correlation functions}
In the local density approximation   
Eqs.(\ref{avdgg}) and  (\ref{CalC2g}) assume the simpler forms
\begin{eqnarray}
\label{dd}
\frac{\delta\beta\Omega_{co}[\rho^*]}{\delta\bar\rho^*({\bf r})}+ \frac{f_h^{(3)}(\bar\rho^*({\bf r}))}{2}{\cal G}_2({\bf r},{\bf r})=0
\end{eqnarray}
and
\begin{eqnarray}
\label{CC}
2{\cal C}_2({\bf r}_1,{\bf r}_2)= 3{\cal C}_2^{co}({\bf r}_1,{\bf r}_2)\\ \nonumber
+
\frac{\beta f^{(4)}_h(\bar \rho^*({\bf r}_1))}{2}\delta({\bf r}_1-{\bf r}_2){\cal G}_2({\bf r}_1,{\bf r}_1)
-\int_{\bf r}\int_{\bf r'}{\cal C}_2^{co}({\bf r}_1,{\bf r}){\cal G}_2({\bf r},{\bf r}'){\cal C}_2^{co}({\bf r}',{\bf r}_2).
\end{eqnarray}
In this approximation Eqs. (\ref{CC}) and (\ref{OZ1}) 
 should be solved selfconsistently. Eq. (\ref{dd}) is the extremum condition for the grand potential. 
\subsection{Periodic structures in the local density approximation - case of weak ordering} 
Let us consider the functional (\ref{DelOM}) for the periodic density profiles (\ref{perdens}), and
let the expansion in $\phi$ be truncated at the fourth order term.
 We restrict our attention to weak ordering,  $\Phi\ll \bar\rho^*_0$.
 For small $\Phi({\bf r})$ the Taylor expansion 
\begin{eqnarray}
\label{Taylor}
 f_h^{(n)}( \bar\rho^*({\bf r}))= f_h^{(n)}( \bar\rho^*_0)+\sum_{m=1}^{\infty}\frac{ f_h^{(n+m)}( \bar\rho^*_0)}{m!}\Phi({\bf r})^m
\end{eqnarray}
can be truncated. In the consistent approximation we truncate the above expansion at the  fourth order term in  $\Phi$. In this approximation Eq.(\ref{DelOM}) assumes for periodic density profiles (\ref{perdens}) the form
\begin{eqnarray}
\label{HaHa11}
 \beta H_{fluc}[\Phi,\phi]=\frac{1}{2}\int_{\bf k}\tilde\phi({\bf k})\tilde C_2^{co}(k)\tilde\phi(-{\bf k})
+\int_{\bf r}
\sum_{n\ge 1}^{'}\frac{{\cal C}^{co}_n[\bar\rho^*_0,\Phi,{\bf r}]}{n!}\phi({\bf r})^n.
\end{eqnarray}
where the prime  in the above sum means that $n\ne 2$,  the explicit expressions for ${\cal C}^{co}_n$ are given in Appendix B, and the function $\tilde C_2^{co}(k)$, defined as in Eq.(\ref{CCga}), has
the explicit form  (see (\ref{co2}))
\begin{eqnarray}
\label{Cco2g}
 \tilde C_2^{co}(k)=\beta\Bigg[\tilde V_{co}(k)+f_h^{(2)}(\bar\rho^*_0)+\frac{f_h^{(4)}(\bar\rho^*_0)}{2}\Phi^2\Bigg]
\end{eqnarray}
with 
\begin{eqnarray}
\Phi^2=\frac{1}{V_u}\int_{{\bf r}\in{\cal V}_u}\Phi({\bf r})^2.
\end{eqnarray}
From (\ref{CC}), (\ref{CCga}), (\ref{GGga}) and (\ref{OZF}) we obtain the equation for $\tilde C_2(k)$
\begin{eqnarray}
\label{C2g}
2\tilde C_2(k)=3\tilde C^{co}_2(k)+\frac{\beta f^{(4)}_h(\bar \rho^*_0)}{2} {\cal G}-\frac{\tilde C^{co}_2(k)^2}{\tilde C_2(k)},
\end{eqnarray}
where we introduced the notation
\begin{eqnarray}
\label{calG}
 {\cal G}=\int_{\bf k}\frac{1}{\tilde C_2(k)}=\int_{\bf k}\tilde G_2(k).
\end{eqnarray}
Recall that by construction of the mesoscopic theory on the length scale $R$, the cutoff $\sim\pi/R$ is present in the above integral. Recall also that  $\int_{\bf k}\tilde G_2(k)=G_2(0)$ \textit{is not the microscopic correlation function at zero distance}, but the correlation function for the {\it microscopic}  density  at the points ${\bf r}'\in S_R({\bf r})$ and ${\bf r}''\in S_R({\bf r})$, \textit{integrated over}  ${\bf r}'\in S_R({\bf r})$ and ${\bf r}''\in S_R({\bf r})$, as discussed in sec.IIIA.
The average density for given $\mu$ and $T$ is the solution of Eq.(\ref{dd}). If there are several solutions, the one corresponding to the global minimum should be chosen. In practice it is much easier to choose the average density $\bar \rho^*_0$ as the independent parameter. 

 In the case of weak ordering, i.e. for $\bar\rho^*({\bf r})$ given in Eqs.(\ref{perdens}) and (\ref{Pp}) with  $\Phi\ll \bar\rho^*_0$,  the truncated  Taylor  expansion of $f_h(\bar\rho^*({\bf r}))$ about $\bar\rho^*({\bf r})=\bar\rho^*_0$ can be inserted in Eq.(\ref{Omcoap}) (see (\ref{Taylor}) for $n=0$), and  $\beta\Omega_{co}[\bar\rho^*]$  can be approximated by
\begin{eqnarray}
\label{OmcoBr}
 \beta\Omega_{co}[\bar\rho^*_0+\Phi]=\beta\Omega_{co}[\bar\rho^*_0] +\frac{1}{2}\int_{\bf k}\tilde\Phi({\bf k})\tilde C_2^{co}(k)\tilde\Phi(-{\bf k})
+
\sum_{n\ge 3}\frac{f_h^{(n)}[\bar\rho^*_0]}{n!}\int_{\bf r}\Phi({\bf r})^n.
\end{eqnarray}
 
The functional (\ref{HaHa11})
can be further simplified in the part of the phase diagram corresponding to the uniform phase, where $\langle \rho^*({\bf r})\rangle=\bar\rho^*_0=const$, just by assuming  $\Phi=0$ in Eqs.(\ref{C1p})-(\ref{C2g}).
Eq.(\ref{dd}) for the uniform phase reduces to the form
\begin{eqnarray}
\label{densa}
 \beta f^{(1)}_h(\bar\rho^*_0)-\mu +\bar\rho^*_0\tilde V_{co}(0)+\beta f^{(3)}_h(\bar\rho^*_0) {\cal G}
=0.
\end{eqnarray}

\subsection{Comparison with the Landau-type theory} 
After all the  assumptions and approximations described in the preceding sections, we finally arrived at the form of $\Omega_{co}$, Eq.(\ref{OmcoBr}), similar to the Landau-Ginzburg-Wilson and Landau-Brazovskii theories. 
In the original Landau-Ginzburg-Wilson and Landau-Brazovskii theories focusing on  universal features of the order-disorder transition,  it is \textit{postulated} that the effective or coarse-grained Hamiltonian in the uniform system has the form
 \begin{equation}
  \label{Landau}
\beta H_{eff}[\phi]=\beta H_{2}[\phi]
+\int_{\bf r}\Big[h\phi({\bf r})+\sum_{n=3}^4\frac{A_n}{n!}\phi^n({\bf r})\Big]
 \end{equation}
where 
\begin{equation}
 \beta H_{2}[\phi]
=\frac{1}{2}\int_{\bf k}\tilde \phi({\bf k})\tilde C^0_2(k)\tilde \phi(-{\bf k}).
\end{equation}
 The summation in (\ref{Landau}) is truncated at $n=6$ when a tricritical point is studied. The form of  $\tilde C^0_2(k)$ is
\begin{equation}
 \label{LGW}
 \tilde C^0_2(k)=\tau_0+\xi_0^2k^2
\end{equation}
or
\begin{equation}
 \label{LB}
\tilde C^0_2(k)=\tau_0+\xi_0^2(k-k_b)^2
\end{equation}
in the LGW and LB theories respectively. In the original Brazovskii theory only even powers of the field are included ($h=A_3=0$). In nonuniform systems, with the equilibrium density profile $\bar\rho^*({\bf r})$, one obtains the effective Hamiltonian $ H_{eff}[\bar\rho^*({\bf r})+\phi({\bf r})]$ as a functional of the fluctuation $\phi({\bf r})$.

Note that (\ref{OmcoBr}) and (\ref{HaHa11}) can be reduced to the LGW or LB functional (\ref{Landau})   with the following expressions for the coupling constants 
\begin{eqnarray}
\label{An}
A_n=\beta f_h^{(n)}(\bar\rho^*_0)
\end{eqnarray}
for $n\ge 3$ and 
\begin{eqnarray}
 h=\beta(f_h^{(1)}(\bar\rho^*_0)-\mu-\tilde V_{co}(0)\bar\rho^*_0)
\end{eqnarray}
in the case of the uniform phase ($\Phi=0$), where the explicit forms of $f_h^{(n)}(\bar\rho^*_0)$  are given in Appendix C for $f_h$ approximated by (\ref{PY}).
However, in our theory (see (\ref{Cco2g}))
\begin{eqnarray}
\label{CbC}
\tilde C^{co}_2(k)=\beta \tilde V_{co}(k)+A_2 +\frac{A_4}{2}\Phi^2.
\end{eqnarray}

 When $\tilde V_{co}(k)$ assumes the global minimum for $k=k_b=0$, and can be expanded about $k=0$, then (\ref{LGW}) corresponds to this expansion truncated at the second  order term.  Truncation of the expansion of $\tilde V_{co}(k)$ is justified when the fields $\tilde\rho^*({\bf k})$ with large $k$ yield negligible contribution to $\Omega$. This is the case when thermally excited density waves with large $ k$ are associated with significantly larger energy than the density waves with the wavenumber $k\to 0$, and can be disregarded. The above conditions are satisfied when the global minimum is deep, and local minima, if exist, correspond to significantly larger values of  $\tilde V_{co}(k)$. More precisely, the interaction potential should satisfy the condition  $|\tilde V_{co}(k)-(\tilde V_{co}(0)+\tilde V_{co}^{(2)}(0)k^2)|/|\tilde V_{co}(k)|\ll 1$ for $k$ that yield the dominant contribution to $\Omega$. Under the above conditions our mesoscopic theory reduces to the Landau-Ginzburg-Wilson theory and describes phase separation. 

When $\tilde V_{co}(k)$ assumes  the global  minimum  $\tilde V_{co}(k_b)<0$ for $k=k_b\ne 0$, and  can be expanded about $k=k_b$, then the truncated expansion 
\begin{eqnarray}
\label{VBV}
\tilde V_{co}(k)=\tilde V_{co}(k_b)+\tilde V_{co}^{(2)}(k_b)(k-k_b)^2/2+...
\end{eqnarray}
 yields $\tilde C_2^{co}(k)$ similar to the LB form (\ref{LB}) (see (\ref{CbC}), and recall that $\Phi=0$ in the uniform phase). Truncation of the expansion of $\tilde V_{co}(k)$ is justified when the fields $\tilde\rho^*({\bf k})$ with large $|{\bf k}-{\bf k}_b|$ yield negligible contribution to $\Omega$. This is the case when the global minimum of $\tilde V_{co}$ is deep, i.e.  the thermally excited density waves with large $|{\bf k}-{\bf k}_b|$ are associated with significantly larger energy than the density waves with the wavenumber $k\approx k_b$. The approximate version of the mesoscopic theory reduces to the Brazovskii theory for such forms of the interaction potentials, and describes microsegregation, or formation of lyotropic liquid crystals. In the following we shall focus on systems with  such forms of $\tilde V_{co}(k)$ - they include weakly charged globular proteins, nanoparticles, colloids or rigid micells in various solvents that mediate effective interaction potentials. Recall that in the mesoscopic theory we  choose the length scale $R$, and consequently introduce the cutoff $\pi/R$. The scale $R$ should be such that the dominant contribution to $\Omega$ comes from the fields $\tilde\rho^*({\bf k})$ with $k<\pi/R$, otherwise  it should be arbitrary. The results of the Landau-Brazovskii theory describe actual ordering when the dominant contribution to $\Omega$  depends on $k_b$, but is independent of $R$. 
\subsection{Brazovskii theory }
Let us briefly summarize the original Brazovskii theory, with the coupling constants expressed in terms of physical quantities according to Eqs.(\ref{An})-(\ref{CbC}) and (\ref{VBV}), and discuss conditions of its validity. The Brazovskii theory is particularly simple, and analytical results can be obtained easily.
In the original Brazovskii theory $\tilde C_2$ is calculated to first order in $A_4$, and is given by \cite{brazovskii:75:0}
\begin{eqnarray}
\label{BrazC}
 \tilde C_2(k)=\tilde C^{co}_2(k)+\frac{A_4}{2}{\cal G}=\tau+\beta^* v_2^*(k-k_b)^2,
\end{eqnarray}
where we simplify the notation by introducing
\begin{eqnarray}
 v_2^*=\frac{\tilde V_{co}^{(2)}(k_b)}{2\tilde V_{co}(k_b)}
\end{eqnarray}
and
\begin{eqnarray}
\label{BrazC1}
 \tau=\tilde C_2(k_b)=\tilde C^{co}_2(k_b)+\frac{A_4}{2}{\cal G}, 
\end{eqnarray}
and where the dimensionless temperature is defined by
\begin{eqnarray}
\label{T*}
 T^*= 1/\beta^*=\frac{k_BT}{-\tilde V_{co}(k_b)}.
 \end{eqnarray}
 $\tilde C^{co}_2(k)$ and ${\cal G}$ are given in Eqs. (\ref{CbC}) and (\ref{calG}) respectively,   and $k_b$ corresponds to the minimum of $\tilde V_{co}$  (see the definition of $V_{co}$ in Eq.(\ref{Vco})).
Here and below $k$ is in $\sigma^{-1}$ units and length is in $\sigma$ units.
Note that (\ref{BrazC}) is consistent with our result (\ref{C2g}) up to a correction which is of order $A_4^2$.
When the fluctuations with
 $k\approx k_b$ dominate,
 then the main contribution to
${\cal G}$ comes from $k\approx k_b$. In
 this case the regularized
integral (\ref{calG}) can be approximated by \cite{brazovskii:75:0,patsahan:07:0}
\begin{equation}
\label{calG1}
{\cal G}=\int_{\bf k} \frac{1}{\tilde C_2(k)}
\simeq\int_{{\bf k}\in S_{\pi/R}} \frac{1}{\tau+\beta^* v_2^*(k-k_b)^2}.
\end{equation}
The integral on the RHS of (\ref{calG1}) can be calculated analytically\cite{patsahan:07:0}. In the case of  $\tau\ll \beta^* v_2^* k_b^2$,  ${\cal G}$ takes the asymptotic form
\begin{equation}
\label{calG0}
{\cal G}\simeq_{\tau\ll \beta^* v_2^* k_b^2} 
{\cal G}(\tau)+\frac{T^*}{2\pi v_2^*R}+O(ln (\pi/R)),
\end{equation}
where
\begin{equation}
\label{calG3}
{\cal G}(\tau)=\frac{2a\sqrt T^*}{\sqrt{\tau}}
\end{equation}
and
\begin{equation}
\label{a}
a=k_b^2/(4\pi\sqrt v_2^*).
\end{equation}
 Note the independence of the  dominant term in (\ref{calG0}) on the mesoscale $R$. In the Brazovskii theory the $R$-dependent terms in (\ref{calG0}) are neglected. The Brazovskii approximation 
\begin{equation}
\label{calG2}
 {\cal G}\simeq{\cal G}(\tau)
\end{equation}
 is  valid when the second term in (\ref{calG0}) is indeed negligible compared to the first term for $1<R<\pi/k_b$, i.e.
\begin{equation}
\label{condval}
 \tau\ll k_b^4\beta^*v_2^*.
\end{equation}
Note also that the original Brazovskii theory is restricted to $k\approx k_b$, because of the approximation (\ref{VBV}).
From (\ref{BrazC1}), (\ref{calG2}) and (\ref{CbC}) we obtain the explicit expression for $\tau=\tilde C_2(k_b)$,
\begin{equation}
\label{r}
 \tau^{3/2}=\tau^{1/2}\tilde C_2^{co}(k_b)+A_4a\sqrt T^*.
\end{equation}

The  grand potential functional (\ref{OmHarval}) in the Brazovskii approximation takes the explicit form 
\begin{eqnarray}
\label{OmHarval2}
\beta\Omega[\bar\rho^*+\Phi]=\beta\Omega_{co}[\bar\rho^*+\Phi]+ 2a\sqrt{\tau T }V-\frac{A_4{\cal G}^2(\tau)}{8}V,
\end{eqnarray}
where 
$\tau$ satisfies Eq.(\ref{r}), $A_2$ and $A_4$ are given in (\ref{A2}) and (\ref{A4}) respectively,  and $\beta\Omega_{co}[\bar\rho^*+\Phi]$ in the Brazovskii-type approximation
 is given in Eq.(\ref{OmcoBr}).  The second term  on the RHS in (\ref{OmHarval2}) is the explicit form of the second term  on the RHS in (\ref{OmHarval})\cite{podneks:96:0,ciach:06:1}. In calculating the third term on the RHS in (\ref{OmHarval}), Eqs.(\ref{BrazC}) and  (\ref{HaHa11}), as well as the property  $\langle \phi({\bf r})^{2n+1}\rangle_G=0$
 were used. The above expression is valid provided that the condition (\ref{condval}) is satisfied.

Global minimum of the functional (\ref{OmHarval2}) with respect to $\Phi$ corresponds to the stable phase for given $\bar\rho^*_0$ and $T^*$. The problem of finding the minimum of (\ref{OmHarval2}) becomes easy for periodic structures of given symmetry. For  given symmetry $\Phi({\bf r})$  can be written in the form
\begin{eqnarray}
\label{Phig}
\Phi({\bf r})=\sum_n\Phi_n g_n({\bf r}),
\end{eqnarray}
where $g_n({\bf r})$ represent the orthonormal basis functions for the $n$-th 
shell that have a particular symmetry, and
satisfy the normalization condition 
\begin{eqnarray}
\label{gnorm}
\frac{1}{V_u}\int_{{\bf r}\in{\cal V}_u}g_n({\bf r})^2=1.
\end{eqnarray}
$\Phi_n$ is the $n$-th amplitude.
For given symmetry the problem reduces to the determination of the minimum of the function of variables $\Phi_n$.
\section{Explicit results}
\subsection{Structure of the disordered phase and the $\lambda$-line}
Let us focus on the stability of the disordered phase. In the first step let us limit ourselves to the MF approximation and consider stability of the functional $\Omega_{co}$. When  $\tilde C^{co}_2(k)<0$, then 
$\Omega_{co}$ is unstable with respect to the density wave with the wavenumber $k$. At the boundary of stability the second functional derivative of $\Omega_{co}$ vanishes for $k=k_b$, i.e.
\begin{eqnarray}
\label{C0}
 \tilde C^{co}_2(k_b)=0,
\end{eqnarray}
since such instability occurs at the highest temperature for given density.
 In the PY approximation the above equation
yields together with (\ref{CbC}) and (\ref{A2}) 
 the \textit{universal} curve 
\begin{eqnarray}
\label{unispin}
 T^*_{\lambda}(\bar\rho_0^*)= \frac{\bar\rho_0^*(1-\eta)^4}{(1+2\eta)^2},
 \end{eqnarray}
 where the dimensionless temperature is defined in (\ref{T*})
and $\tilde V_{co}(k)$ assumes the global minimum, $\tilde V_{co}(k_b)<0$, for $k=k_b$. 
Universality in this context means that the shape of $\tilde V_{co}(k)$ is irrelevant, and the value at the minimum, $\tilde V_{co}(k_b)$, sets the temperature scale. For properly rescaled temperatures the boundaries of stability of $\Omega_{co}$ for all systems with particles having spherical cores collaps onto the single master curve (\ref{unispin}).

For $k_b=0$, Eq.(\ref{unispin})  represents the MF approximation for the spinodal line of the gas-liquid separation, whereas for $k_b\ne 0$ the above represents the $\lambda$-line \cite{stell:99:0,ciach:00:0,ciach:01:1,ciach:03:1,ciach:03:0,ciach:07:0,patsahan:07:0,archer:07:0,archer:07:1} associated with microsegregation. Similar result was obtained previously \cite{sear:99:0,archer:07:0,archer:07:1,archer:08:0,archer:08:1}.  The universal line (\ref{unispin}) is shown in Fig.3 (dashed line). 

Let us consider the actual boundary of stability of the uniform phase in the system in which $\tilde V_{co}(k)$ assumes the global minimum for $k_b>0$. Beyond MF the boundary of stability  of the grand potential $\Omega$ is given by
 \begin{eqnarray}
\label{ist}
 \tilde C_2(k_b)=0.
\end{eqnarray}
In the Brazovskii theory  $\tilde C_2(k_b)=\tau$ satisfies the equation (\ref{r}). Eqs. (\ref{ist}) and  (\ref{r}) yield 
$ T=0$ independently of density.
This means that when $\tilde V_{co}(k)$ assumes the global minimum for $k_b>0$, the MF boundary of stability with respect to periodic ordering, Eq.(\ref{unispin}), is shifted down to zero temperature when the mesoscopic scale fluctuations are included. 
At the $\lambda$-line the structure factor $S(k)=\tilde G_2(k)/\bar\rho^*_0+1$ diverges for $k\to k_b$ for $\tilde G_2(k)$ approximated by $\tilde G^{co}_2(k)=1/\tilde C^{co}_2(k)$. In the Brazovskii approximation the maximum of the structure factor is  finite at the  $\lambda$-line,  and we obtain its value from (\ref{r}) and (\ref{C0}),
\begin{eqnarray}
\label{Gla}
 \tilde G_2(k_b)=1/\tau=\Bigg[\frac{(4\pi)^2v_2^*A_2}{k_b^4A_4^2}\Bigg]^{1/3}.
\end{eqnarray}
Recall that (\ref{r}) is valid provided that $\tau$  satisfies the condition (\ref{condval}), which at the $\lambda$ line takes the form
\begin{eqnarray}
\label{ccoco}
 \frac{A_4}{ A_2^2}\ll 4\pi (v_2^*k_b^2)^2.
\end{eqnarray}
 For $\bar\rho^*_0\to 0$ the approximation (\ref{calG2}) is not valid at the $\lambda$-line, because the LHS of (\ref{ccoco})  behaves as $(\bar\rho^*_0)^{-1}$ (see Appendix C), whereas the RHS   is independent of  $\bar\rho^*_0$. The  RHS of (\ref{ccoco}) depends on the system, therefore the density range for which the approximation (\ref{calG2}) is valid at the $\lambda$-line, is  system-dependent. The condition (\ref{condval}) is satisfied  below the $\lambda$-line for sufficiently low $T^*$ (see (\ref{r}) for $\tilde C_{co}(k_b)<0$). The first-order transitions in the Brazovskii theory occur on the low-temperature side of the $\lambda$-line \cite{brazovskii:75:0}, therefore the simple approximate theory can be applied. 

The infinite susceptibility with respect to periodic external field, resulting from singularity of the structure factor for finite $k$,  is the artifact of the MF approximation. 
In the mesoscopic theory the interpretation of the $\lambda$ line follows from Eqs.(\ref{p2}) and (\ref{OmcoBr}), which show that   when  $\tilde C^{co}_2(k_b)<0$, the mesoscopic density $\bar\rho^*({\bf r})=\bar\rho_0^*+\Phi g_1({\bf r})$, with infinitesimal-amplitude $\Phi$ and $g_1$ given in Appendix D, is more probable than the mesoscopic density $\bar\rho^*({\bf r})=\bar\rho_0^*$. This is because for infinitesimal $\Phi$ the remaining contributions in Eq.(\ref{OmcoBr}) are irrelevant. 
 For $T^*<T^*_{\lambda}(\bar\rho_0)$ (i.e. $\tilde C^{co}_2(k_b)<0$) the \textit{single mesostate with periodic density},  with the wavenumber $k_b$ and infinitesimal amplitude, \textit{is more probable} than the \textit{uniform mesostate}. Recall that the mesoscopic density $\rho({\bf r})$ is equivalent to the set of microstates that satisfy (\ref{mesor}). Thus, for $T^*<T^*(\bar\rho_0)$ the probability of occurrence of \textit{any microscopic state with nonuniform density distribution} at the length scale $\pi/k_b$  is larger than probability of finding  \textit{any microscopic state with position-independent density} (\ref{mesor}) on the length scale $R\sim\pi/k_b$.  However, averaging over all microscopic states with different  spatial inhomogeneities may lead to space independent average density, unless the system undergoes a\textit{ first-order transition }to ordered phases with \textit{finite amplitude} $\Phi$ of density oscillations. The first-order transition can be determined beyond the stability analysis.

When  $\tilde V_{co}(0)<0$, then both, the $\lambda$-line and the spinodal line are present and the temperature at the MF spinodal line is
\begin{eqnarray}
T^*_s(\bar\rho_0^*)=T^*_{\lambda}(\bar\rho_0^*) \frac{\tilde V_{co}(0)}{\tilde V_{co}(k_b)}. 
 \end{eqnarray}
$T^*_s(\bar\rho_0^*)<T^*_{\lambda}(\bar\rho_0^*)$ when $\tilde V_{co}(k)$ assumes the global minimum for $k_b>0$. Both lines were found in Ref.\cite{archer:07:0,archer:07:1,archer:08:1} for some forms of the interaction potential. In fact there is a family of curves $T^*_{\lambda}(\bar\rho_0^*)\tilde V_{co}(k)/\tilde V_{co}(k_b)$ representing the MF instability with respect to density waves with the wavenumber $k$. All of them lie below the $\lambda$-line when $k_b>0$, or below the spinodal line when $k_b=0$. 
\begin{figure}
\includegraphics[scale=0.35]{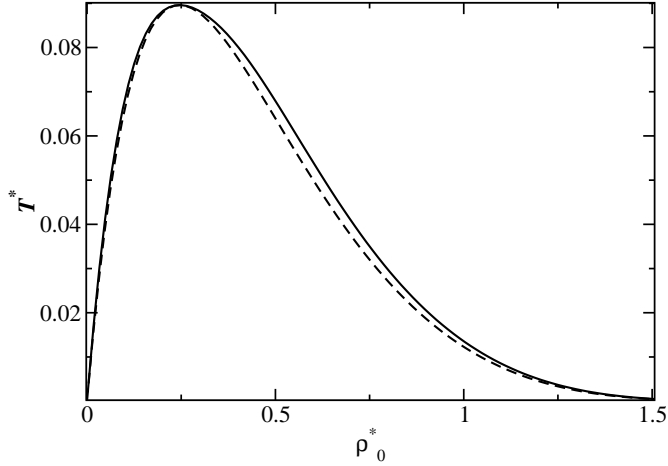}
\caption{Dashed line is the universal line (\ref{unispin}) that represents the MF spinodal line of the gas-liquid separation when $\tilde V_{co}(k)$ assumes the global minimum for $k=0$, or the  $\lambda$-line associated with formation of ordered periodic phases when $\tilde V_{co}(k)$ assumes the global minimum for $k=k_b>0$. Solid line is the universal first-order transition  line between the disordered and the bcc phases in the one-shell MF approximation.
 $T^*$ and $\rho^*_0$ are dimensionless temperature and density defined in Eqs.(\ref{T*}) and (\ref{dimless}) respectively.}
\end{figure}

\subsection{Case of weak ordering - universal features of the MF phase diagram}
Let us determine the most probable mesoscopic states for given thermodynamic variables. 
The most probable mesostate corresponds to the global minimum of the functional $\beta\Omega_{co}[\bar\rho^*_0+\Phi]$ with respect to  $\Phi({\bf r})$ - the position dependent density deviation from the mean value $\bar\rho^*_0$ (see (\ref{p})).  On the other hand, the minimum of $\Omega_{co}$ is equivalent to the MF approximation, where the second contribution to the grand potential $\Omega$ in Eq.(\ref{FF}) is neglected. 
The minimum of the functional of the same formal structure as in Eq. (\ref{OmcoBr}) was calculated in different context in Ref.\cite{brazovskii:75:0,leibler:80:0,fredrickson:87:0,podneks:96:0}, and we shall not repeat the details of the calculation which can be found in the above papers.

In this section we shall limit ourselves to the one-shell approximation. This approximation is valid when  $\tilde V_{co}(k_b) \ll \tilde V_{co}(k_{b2})$, where $k_{b2}$ is the wavenumber in the second shell.
For structures possesing different symmetries the Fourier transform of
$g_1$ in (\ref{Phig}) has the form
\begin{equation}
\label{g}
\tilde g_1({\bf k})=\frac{ (2\pi)^{d}}{\sqrt{2n}}
\sum_{j=1}^{n}\Big(w\delta({\bf k}-{\bf k}^j_{b})+
w^*\delta({\bf k}+{\bf k}^j_{b})\Big),
\end{equation}
where $w^*$ is the complex conjugate of $w$, and $ww^*=1$. $2n$ is the number of vectors ${\bf k}^j_{b}$ in the first
shell of the considered structure. The forms of $g_1({\bf r})$ in real-space representation are given in Appendix D for the lamellar,  hexagonal,  bcc  and gyroid (Ia3d) structures. 
 From (\ref{OmcoBr}) 
we obtain in the one-shell
approximation
\begin{eqnarray}
\label{OO}
\beta\Delta\omega_{co}(\Phi_1)=\beta\Big(\Omega_{co}[\bar\rho^*+\Phi_1g_1({\bf r})]-\Omega_{co}[\bar\rho^*]\Big)/V=\frac{1}{2}\tilde C_2(k_b)\Phi_1^2+
\frac{A_3\kappa_3}{3!}\Phi_1^3
+\frac{A_4\kappa_4}{4!}\Phi_1^4,
 \end{eqnarray}
where the geometric factors characterizing different structures are given by
\begin{eqnarray}
\label{kappa}
\kappa_n=\frac{1}{V_u}\int_{{\bf r}\in {\cal V}_u}g_1({\bf r})^n.
\end{eqnarray}
Note that in the one-shell approximation $\beta\Delta\omega_{co}(\Phi_1)$ depends on $\tilde V_{co}(k)$  only through the product $\beta\tilde V_{co}(k_b)$ (see (\ref{CbC})), therefore the phase diagram in variables $\bar\rho^*_0, T^*$ is universal.
From the extremum condition
\begin{eqnarray}
\label{OOO}
\frac{\partial \beta\Delta\omega_{co}}{\partial \Phi_1}=
 \tilde C_2(k_b)\Phi_1+\frac{A_3\kappa_3}{2}\Phi_1^{2}+
\frac{A_4\kappa_4}{3!}\Phi_1^{3}=0
\end{eqnarray}
we obtain $\Phi_1$ of the  stable or the  metastable phase. The stable  phase  corresponds to the lowest value of  $\beta\Delta\omega_{co}(\Phi_1)$ for given thermodynamic variables, where $\Phi_1$ satisfies (\ref{OOO}). The coexistence between different ordered phases takes place when the grand potentials (\ref{OO}) for these phases are equal. 
At the  coexistence of the stable ordered phase with the
disordered (uniform) phase  
\begin{eqnarray}
\label{eqod}
 \Delta\omega_{co}(\Phi_1)=0.
\end{eqnarray}
The universal MF phase diagram obtained in this way in the 
 one-shell approximation is shown in Fig. 4.
For $A_3\ne 0$ the
ordered phase coexisting with the fluid has the bcc symmetry, 
and the uniform-bcc phase coexistence line is 
\begin{eqnarray}
\label{trli}
T^*=\frac{3A_4\kappa_4^{bcc}}
{3A_4A_2\kappa_4^{bcc}-(A_3\kappa_3^{bcc})^2}
\end{eqnarray}
 as already shown by Leibler \cite{leibler:80:0}. Explicit expressions for $A_n$ are given in Appendix C, and the transition line (\ref{trli}) is shown in Fig.3 together with the $\lambda$-line, and in Fig. 4 together with transitions to the other phases.  
The lattice constant of the bcc phase is $a=2\sqrt 2\pi/k_b$. 
Along the coexistence between the uniform and the bcc phases 
\begin{eqnarray}
\label{PP}
\Phi_1^{bcc}=-\frac{16A_3}{15\sqrt 3A_4},
\end{eqnarray}
where the explicit forms of the geometric coefficients, $\kappa_3^{bcc}=2/\sqrt 3 $ and $\kappa_4^{bcc}=15/4$, were used.
$\Phi_1^{bcc}$ at the transition to the uniform phase is independent of $V_{co}$, i.e. in the MF one-shell approximation is universal,
 and is shown in Fig.5 for our PY form of $f_h$.  For $A_3=0$ the
transition is to the striped (lamellar) phase\cite{leibler:80:0} and is continuous $\Phi_1=0$). In our PY theory $A_3(\rho^*)=0$ for  just one density $ \rho^*=\rho^*_c$, where $ \rho^*_c$ is the maximum at the $\lambda$-line, and at the same time the critical density of the gas-liquid
separation. In the PY approximation (\ref{PY})  $\rho^*_c\approx 0.2457358$.  Note that for
 densities lower than the gas-liquid critical-point density $\rho^*_c$ we obtain
 a periodic array of excess number density (cluster) forming the bcc
 crystal, whereas for higher densities the bcc structure is formed by
 bubbles of depleted density. Note also that $\Phi_1\ll 1$ as required by the construction of the approximate theory (see sec.IIIB).
The difference in the average densities of the coexisting phases  in this approximation vanishes. 

The diagram shown in Fig.4 is the universal 'skeleton' showing the sequence of phases: disordered, bcc, hexagonal, lamellar, inverted hexagonal, inverted bcc, disordered. Note the re-entrant melting at densities well below the close-packing density. The above sequence of phases agrees with the sequence observed for micellar solutions and block copolymers, and  with the  sequence of structures found in recent simulations for the SALR potential \cite{archer:07:1,candia:06:0}. However, bicontinuous cubic phases found in some of the self-assembling systems are not present on the universal MF diagram obtained in the one-shell approximation.

\begin{figure}
\includegraphics[scale=0.35]{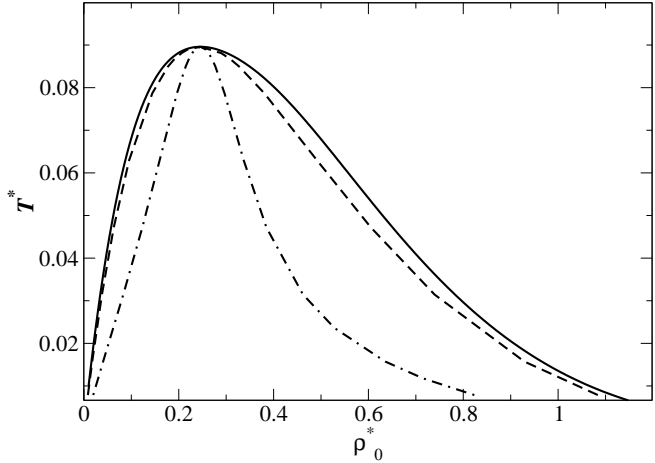}
\caption{MF phase diagram in the one-shell approximation. $T^*$ and $\rho^*_0$ are dimensionless temperature and density, Eqs.(\ref{T*}) and (\ref{dimless}) respectively. Solid line is the coexistence of the uniform phase (above in $T^*$) and the bcc crystal. On the left from the maximum the bcc structure is formed by droplets (excess density), on the right by bubbles (depleted density). The bcc crystal coexists with the hexagonal structure along the dashed line. Again, on the left and on the right from the maximum  the  hexagonally packed cylinders consist of droplets and bubbles respectively. The hexagonal phase coexists with the lamellar phase along the dash-dotted line. 
}
\end{figure}

\begin{figure}
\includegraphics[scale=0.35]{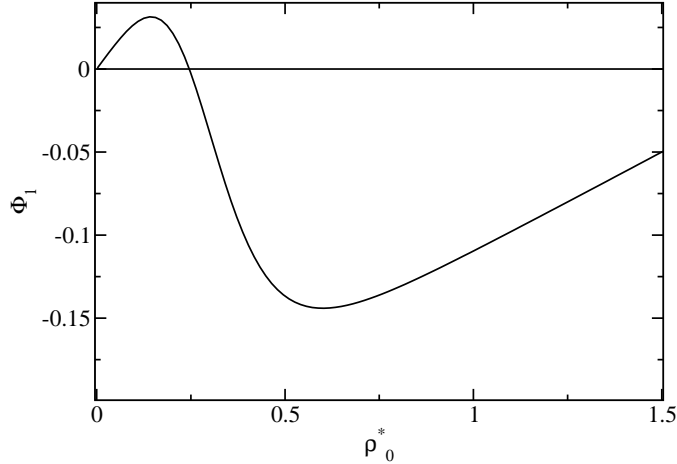}
\caption{The amplitude of the density profile $\bar\rho^*_0({\bf r})=\bar\rho^*_0+\Phi_1g_1^{bcc}({\bf r})$ in the bcc phase at the coexistence with the uniform phase (i.e., along the solid line in Fig.4) in the one-shell MF approximation. $\Phi_1$ and the space-averaged density $\bar\rho^*_0$ are both dimensionless. 
}
\end{figure}
\subsection{Example: the SALR potential in the case of very short range of attractions}
 Beyond the one-shell approximation more complex structures can be stabilized, and we expect 'decorations' of the universal skeleton diagram (Fig.4) with regions of stability of more complex structures, or with structures with large amplitudes of the density oscillations. However, the diagrams are no longer universal, in the sense that the stability region of more complex structures depends not only on the value of $\beta\tilde V_{co}(k)$ at the minimum at $k=k_b$, but also on  $\beta\tilde V_{co}(k_{bn})$, where $k_{bn}$ is the wave number in the $n$-th shell. Therefore the details of the phase diagram depend on the shape of the interaction potential.
Studies of the details of the phase diagrams in various systems  go beyond the scope of this work. Just for illustration we quote the results obtained for the SALR potential
\begin{equation}
\label{SALR}
V_{SALR}(r)=-A_a\frac{\exp(-z_1r)}{r}+A_r\frac{\exp(-z_2r)}{r}
\end{equation}
where $r$ and $z_i$ are  in $\sigma$ and $1/\sigma$ units respectively, and the amplitudes are in $k_BT_{room}$ units. 
The SALR potential 
describes in particular  weakly charged colloids in a presence of short-chain non-adsorbing polymers inducing the depletion potential, globular proteins in some solvents, and rigid micells. For the parameters $A_a=140e^{8.4}$, $A_r=30e^{1.55}$, $z_1=8.4$ and $z_2=1.55$, related to the colloid-polymer mixture \cite{campbell:05:0}, the Fourier transform of the potential $ V_{co}(r)=\theta(r-1)V_{SALR}(r)$ is shown in Fig.6 together with the approximation (\ref{VBV}) that allows for the reduction to the Brazovskii theory. This potential leads to formation of small clusters, because $k_b\approx 1.94$ in $\sigma^{-1}$ units, and $2\pi/k_b\approx 3.24$. In Fig. 7 $\tilde G_2^{co}(k)=(S_{co}(k)-1)\bar\rho^{*}_0$, where $S_{co}(k)$ is the structure factor in MF  approximation, is shown for $\bar\rho^*_0=\bar\rho^*_c\approx0.246$ and $T^*=0.15$. The first peak corresponds to cluster-cluster correlations, as observed in simulations and experiments for the SALR systems \cite{archer:07:1,sciortino:04:0,sciortino:05:0,campbell:05:0,stradner:04:0}. The chosen thermodynamic state is away from the 
$\lambda$-line, and  $\tilde G_2^{co}(k)$  is a reasonable approximation for the correlation function. 
When the $\lambda$-line is approached, $\tilde G_2^{co}(k_b)$ diverges and the MF approximation fails, as discussed in sec.Va. In Fig.8  $\tilde G_2(k_b)$ (maximum of the structure factor) is shown  for $\bar\rho^*_0=\bar\rho^*_c\approx0.246$ as a function of temperature for 
$T^*\le T^*_{\lambda}( \bar\rho^*_c)$ in the Brazovskii approximation. In this case the condition (\ref{condval}) is satisfied. We verified that inclusion of the second term in Eq.(\ref{calG0}) with $R=1$ and $R=\pi/k_b$ leads to $\tilde G_2(k_b)\approx 0.14$ and $\tilde G_2(k_b)\approx 0.15$ respectively at the $\lambda$-line, while $\tilde G_2(k_b)\approx 0.16$ when the second term in (\ref{calG0}) is neglected. Note that $\tilde G_2(k_b)$ assumes rather small values, in agreement with results obtained for ionic systems \cite{patsahan:07:0}.

\begin{figure}
\includegraphics[scale=0.45]{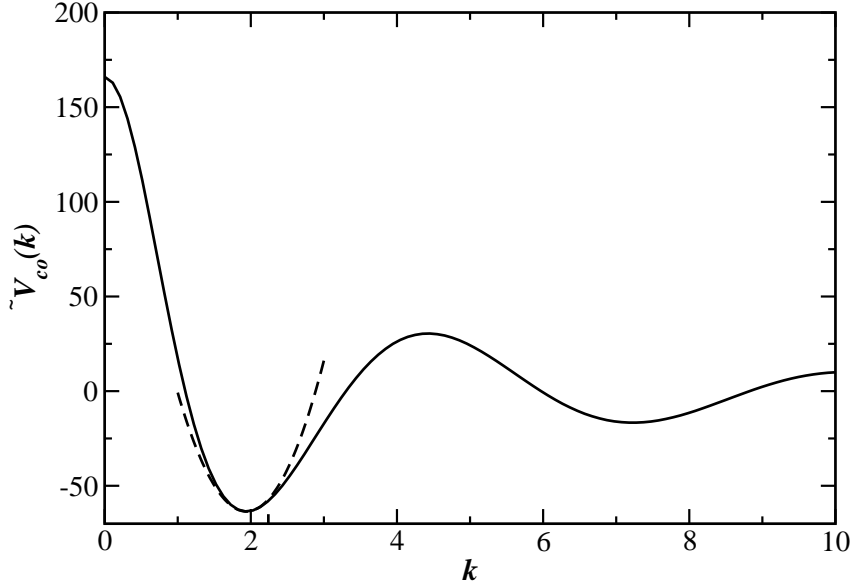}
\caption{The Fourier transform $\tilde V_{co}(k)$  of $V_{co}(r)=\theta(r-1)V_{SALR}(r)$, where $ V_{SALR}(r)$ is the SALR potential (\ref{SALR}) with $A_a=140e^{8.4}$, $A_r=30e^{1.55}$, $z_1=8.4$ and $z_2=1.55$ (solid line). Dashed line represents Eq. (\ref{VBV}) that leads to the approximate Brazovskii theory.   The minimum of $\tilde V_{co}(k)$ is assumed at $k=k_b\approx 1.94$. The second-shell value $k_{b2}=2k_b/\sqrt 3$ is indicated on the $k$-axis.  $\tilde V_{co}(k)$ and $k$ are in $k_BT_{room}$ and $1/\sigma$ units respectively. Brazovskii theory is valid when the dominant deviations from uniform distribution of particles correspond to $k\approx k_b$. }
\end{figure}

\begin{figure}
\includegraphics[scale=0.35]{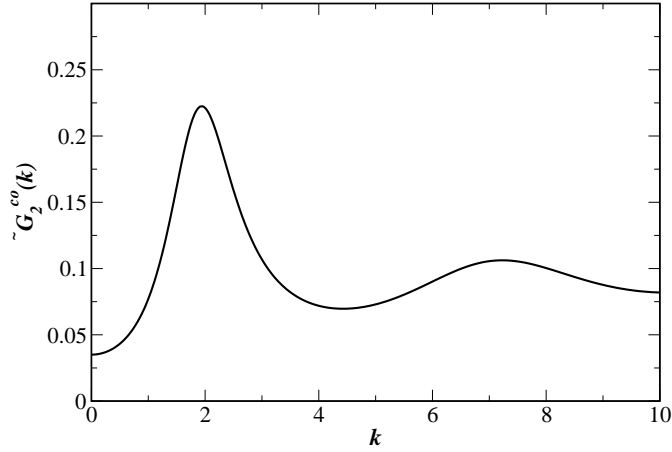}
\caption{ $\tilde G_2^{co}(k)=(S_{co}(k)-1)\bar\rho^{*}_0$ for the SALR potential  (\ref{SALR}) with $A_a=140e^{8.4}$, $A_r=30e^{1.55}$, $z_1=8.4$ and $z_2=1.55$  for $\bar\rho^*_0=\bar\rho^*_c\approx 0.246$ and $T^*=0.15$.
}
\end{figure}
\begin{figure}
\includegraphics[scale=0.35]{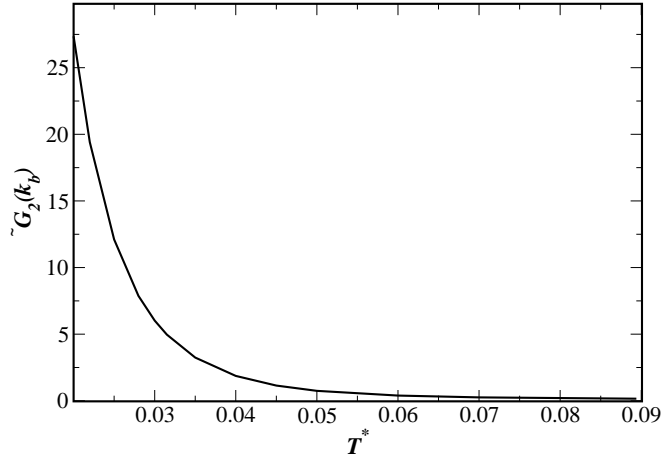}
\caption{ $\tilde G_2(k_b)=1/\tau$, corresponding to the maximum of the structure factor,  in the Brazovskii approximation (\ref{r}), for $\bar\rho^*_0=\bar\rho^*_c\approx 0.246$ as a function of $T^*$. }
\end{figure}
We find the diagram shown in Fig.9 by calculating $\beta\Delta\omega_{co}$ for the gyroid phase in the two-shell approximation.  In this phase the second shell is very close to the first shell ($k_{b2}=2k_b/\sqrt 3$), unlike in the case of the other phases ($k_{b2}=\sqrt 2k_b$, $k_{b2}=\sqrt 3k_b$ and $k_{b2}=2k_b$ for the bcc, hexagonal and lamellar phase respectively).  For this reason  the second shell in (\ref{Phig}) should be included in the case of the gyroid phase, as argued in Ref.\cite{podneks:96:0}. The details of the calculation will be given elsewhere along with the results obtained within the present mesoscopic theory beyond MF\cite{ciach:08:2}.
The unit cell of the gyroid phase  is shown in Fig.10 for $\bar\rho^*_0=0.048$ and $T^*=0.016$. Note that in this approximation we obtain for low enough temperatures the sequence of phases: disordered-bcc-hexagonal-gyroid-lamellar-inverted gyroid-inverted hexagonal -inverted bcc-disordered. Such sequence of phases is found in many self-assembling systems, including micellar systems and block copolymers. The bicontinuous cubic phase is usually found between the hexagonal and the lamellar phases. In simulations of the SALR potential   hexagonal and lamellar phases were found for similar densities as we predict\cite{candia:06:0}. The simulations were restricted to $\bar\rho^*<0.25$, therefore the inverted structures were not found. However, for the form of the SALR potential corresponding to much larger $\pi/k_b$, spherical, cylindrical and slab-like liquid-like clusters were found for increasing density, and  for  $\bar\rho^*>0.35$ cylindrical and next spherical bubbles were seen  \cite{archer:07:1} for densities that  agree with our predictions.
The unit cell in this system contains too many particles to enable observation of the ordered phases. In colloid-polymer mixtures spherical clusters, elongated clusters and a network forming cluster were observed experimentally for increasing volume fraction of colloids \cite{campbell:05:0}.  Relation between the experimentally observed (presumably metastable) structures and our results obtained for thermal equilibrium requires further studies.   
\begin{figure}
\includegraphics[scale=0.35]{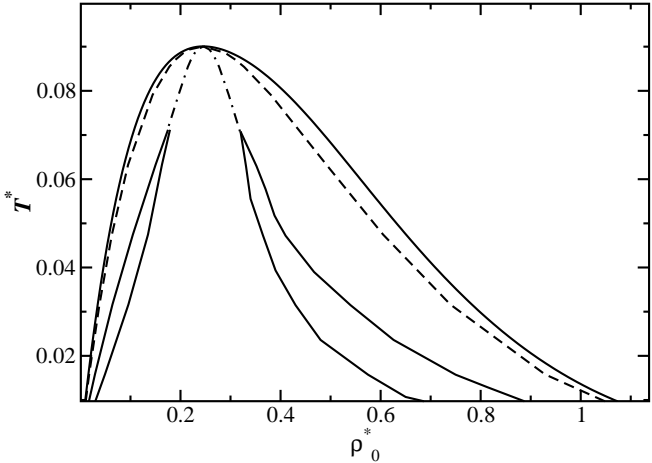}
\caption{MF phase diagram with the gyroid phase  (Ia3d symmetry) considered in the two-shell approximation, and the remaining phases in the one-shell approximation,  as in  Ref.\cite{podneks:96:0}. The interaction potential has the SALR form (\ref{SALR}) with $A_a=140e^{8.4},z_1=8.4,A_r=30e^{1.55},z_2=1.5$.  The potential is chosen for the system consisting of charged colloids in the presence of small polymers, similar to the system studied experimentally in Ref.\cite{campbell:05:0}. The diagram is the same as in Fig.5 except that the Ia3d phase is stable in the windows between the hexagonal and the lamellar phases. The structure of the phase in the left window is shown in Fig.10. For more details see Ref.\cite{ciach:08:2}.
}
\end{figure}
\begin{figure}
\includegraphics[scale=0.35]{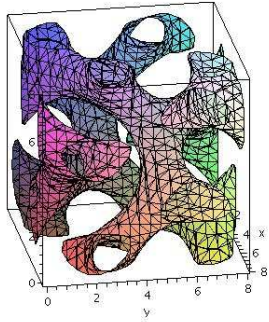}
\caption{ The unit cell of the Ia3d phase  in the two shell approximation for $\bar\rho^*_0=0.048$ and $T^*=0.016$. In the region enclosed by the shown surface $\bar\rho^*({\bf r})-\bar\rho^*_0\ge 0.06$. Note that two networks of branching regions of excess density are present in the two-shell approximation in the structure corresponding to the minimum of $\beta\Delta\omega_{co}$. The lattice constant of the unit cell is $2\pi\sqrt 6/k_b\approx 8\sigma$. For more details see Ref.\cite{ciach:08:2}.
}
\end{figure}

The effect of the fluctuation contribution (the two last terms in Eq.(\ref{OmHarval2})) on the phase equilibria will be described in Ref.\cite{ciach:08:2}, where more details on the two-shell approximation will also be given. In short - when the fluctuation contribution in Eq.(\ref{OmHarval2}) is included, the stability region of the disordered phase enlarges, and the effect increases with increasing temperature.  For low temperatures the sequence of phases is not affected by the mesoscopic fluctuations.

\section{Final remarks}

We developed a formalism that has a form of the  density functional  theory with additional term in which the effect of the mesoscopic scale fluctuations is included. The additional term depends on the correlation function for the density-fluctuation correlations on the mesoscopic length scale, for which self-consistent equations are derived. The fluctuation contribution can be obtained when the form of the grand potential with frozen mesoscopic scale fluctuations is known from the microscopic theory. 
The theory is designed for studies of self-assembling systems, where nonuniform density distributions are found on the mesoscopic length scale, and allows for obtaining phase diagrams and structure  in terms of density, temperature and the parameters characterizing (effective) interaction potentials, as well as microscopic correlation function. We present the grand-potential functional and  the equations for the correlation functions, starting from the most general  form (Eq.(\ref{FF})). In sec.IV we describe approximate version of the theory that allows for analytical calculations. The local density approximation was recently compared with more accurate Rosenfeld density functional for the SALR potential, and good agreement was obtained for not too high densities~\cite{archer:08:1}. We also show that our theory can be reduced to either the LGW or the LB theory. The range of validity of the latter approximate theories is discussed in detail. Explicit  expressions for the coupling constants as well as the form of the grand potential and the correlation function are given for the Brazovskii approximation. The above mentioned equations can be applied to a wide class of interaction potentials describing various soft-matter systems with particles having spherical cores. The theory can be generalized to nonspherical cores.  

In sec.V  universal properties of selfassembly are determined in the framework of the mesoscopic theory introduced in sec. II-IV. In addition, a particular example of the SALR potential is considered. 
We first determine the  $\lambda$-line and conclude that it is related to structural changes in the disordered system. On the low-temperature side of the $\lambda$-line spatial inhomogeneities of a particular size $\lambda$ appear more frequently than densities homogeneous in regions of linear dimension $\lambda$. 
 We obtain universal 'skeleton' diagram (Fig.4) with the universal sequence of ordered phases that agrees with many experimental observations in a wide class of self-assembling systems, and with simulation studies for the SALR potential\cite{archer:07:1,candia:06:0}. This universal skeleton is 'decorated' in particular systems with more complex structures, and modified by fluctuations, especially for high temperatures. 
The continuous transition to the lamellar phase obtained in MF becomes fluctuation-induced first order when the mesoscopic scale fluctuations are included. The density difference  between the disordered and the lamellar  phase and between different ordered phases vanishes on this level of approximation. The transitions are presumably very weakly first order, therefore it is difficult to distinguish them from the continuous transitions in simulations \cite{archer:07:1}. 

 By considering the  gyroid phase  in the two-shell approximation for a particular form of the SALR potential, we find that this phase is stable between the hexagonal and lamellar phases for low enough temperatures (Fig.9). The bicontinuous phase can be considered as a regular gel. Relation between this phase and the experimentally observed gels in colloid -polymer mixtures \cite{campbell:05:0} is an interesting question. If the gyroid phase is  thermodynamically stable, the gel observed 
 in experiments  may result from arrested microsegregation, and its structure should be more regular than the structure that is formed by arrested spinodal decomposition into two uniform phases. In particular, triple junctions of the cylindrical colloidal clusters should dominate.

{\bf Acknowledgments}
I would like to thank Dr. A. Archer for discussions and for sending the preprints (Refs. \cite{archer:08:0} and \cite{archer:08:1}). This work was partially supported by the Polish Ministry of Science and Higher Education, Grant No NN 202 006034. 

\section{appendices}
\subsection{Two-point vertex function in Fourier representation in the case of periodic structures}
For periodic structures with the unit cell ${\cal V}_u$ of the volume $V_u$, with the structure invariant with respect to translations by the vector ${\bf P}=\sum_i^3n_i{\bf p}_i$, where $n_i$ are integer and the vectors ${\bf p}_i$ span the unit cell, we have
\begin{eqnarray}
 \tilde {\cal C}_2({\bf k},{\bf k}+{\bf k}')=\int_{{\bf r}_1}\int_{{\bf r}_2}{\cal C}_2({\bf r}_1-{\bf r}_2|{\bf r}_1 )e^{i{\bf k}\cdot ({\bf r}_1-{\bf r}_2)}e^{i({\bf k}+{\bf k}')\cdot {\bf r}_2}=\\
\nonumber 
V_u\sum_{\bf P}e^{i({\bf k}+{\bf k}')\cdot {\bf P}}\frac{1}{V_u}\int_{{\bf r}_2\in{\cal V}_u}e^{i({\bf k}+{\bf k}')\cdot {\bf r}_2} \int_{{\bf r}_1-{\bf r}_2}e^{i{\bf k}\cdot ({\bf r}_1-{\bf r}_2)}{\cal C}_2({\bf r}_1-{\bf r}_2|{\bf r}_2)=
\\
\nonumber \delta({\bf k}+{\bf k}')\int_{\bf r_1-r_2}e^{i{\bf k}\cdot({\bf r_1-r_2})}\frac{1}{V_u}\int_{{\bf r}_2\in{\cal V}_u}{\cal C}_2({\bf r}_1-{\bf r}_2|{\bf r}_2)
\end{eqnarray}

\subsection{Explicit forms of ${\cal C}^{co}_n$ for periodic structures in the local density approximation}
From (\ref{CoPY}) we obtain
\begin{eqnarray}
\label{C1p}
 {\cal C}^{co}_1[\bar\rho^*_0,\Phi,{\bf r}_1]=\beta\Big[f_h^{(1)}(\bar\rho^*_0)+f_h^{(2)}(\bar\rho^*_0)\Phi({\bf r}_1)+\frac{f_h^{(3)}(\bar\rho^*_0)}{2}\Phi({\bf r}_1)^2+\frac{f_h^{(4)}(\bar\rho^*_0)}{3!}\Phi({\bf r}_1)^3
\\
\nonumber
-\mu+\int_{\bf r_2}V_{co}(r_{12})(\bar\rho^*_0+\Phi({\bf r}_2))\Big]
\end{eqnarray}
\begin{eqnarray}
\label{co2}
{\cal C}^{co}_2[\bar\rho^*_0,\Phi,{\bf r}_1,{\bf r}_2]=\beta\Bigg[ \Bigg(f_h^{(2)}(\bar\rho^*_0)+f_h^{(3)}(\bar\rho^*_0)\Phi({\bf r}_1)+\frac{f_h^{(4)}(\bar\rho^*_0)}{2}\Phi({\bf r}_1)^2\Bigg)\delta({\bf r}_1-{\bf r}_2)+V_{co}(r_{12})\Bigg]
\end{eqnarray}
\begin{eqnarray}
 {\cal C}^{co}_3[\bar\rho^*_0,\Phi,{\bf r}]=\beta\Big[f_h^{(3)}(\bar\rho^*_0)+f_h^{(4)}(\bar\rho^*_0)\Phi({\bf r})\Big]
\end{eqnarray}
\begin{eqnarray}
\label{C4l}
 {\cal C}^{co}_4[\bar\rho^*,\Phi]=\beta f_h^{(4)}(\bar\rho^*_0).
\end{eqnarray}

\subsection{Expressions for the coupling constants $A_n$ in the PY approximation}
The explicit forms of $A_n=f_h^{(n)}(\bar\rho^*_0)$, for $f_h$ approximated by (\ref{PY}) are
\begin{eqnarray}
\label{A2}
A_2=\frac{(1+2\eta)^2}{(1-\eta)^4\bar\rho_0^*}
\end{eqnarray}
\begin{eqnarray}
\label{A3}
A_3=\frac{12\eta^3+20\eta^2+5\eta-1}{(1-\eta)^5\bar\rho_0^{*2}}
\end{eqnarray}
\begin{eqnarray}
\label{A4}
A_4=\frac{2-12\eta+30\eta^2+112\eta^3+48\eta^4}{(1-\eta)^6\bar\rho_0^{*3}}
\end{eqnarray}
\subsection{$g_1({\bf r})$ for several structures}
The function $g_1({\bf r})$ is given for the lamellar, hexagonal, bcc and gyroid phases in Eqs.(\ref{lam})-(\ref{giro}) respectively, with ${\bf r}=(x_1,x_2,x_3)$. The minimum of $\tilde V_{co}(k)$ corresponds to $k_b$.
\begin{eqnarray}
\label{lam}
g^{\ell}_1({\bf r})=\sqrt 2\cos(k_b x_1)
\end{eqnarray}
\begin{eqnarray}
\label{hex}
g^{hex}_1({\bf r})=\sqrt{\frac{2}{3}}\Bigg[\cos(k_bx_1)+
2\cos\Big(\frac{k_bx_1}{2}\Big)\cos\Big(\frac{\sqrt 3 k_bx_2}{2}\Big)\Bigg]
\end{eqnarray}
\begin{eqnarray}
\label{bbcc}
g^{bcc}_1({\bf r})=\frac{1}{\sqrt 3}\sum_{i<j}
\Bigg(\cos\Big(\frac{k_b(x_i+x_j)}{\sqrt 2}\Big)
+\cos\Big(\frac{k_b(x_i-x_j)}{\sqrt 2}\Big)\Bigg)
 \end{eqnarray}
\begin{eqnarray}
\label{giro}
g^{giro}_1({\bf r})=\sqrt{\frac{8}{3}}\Bigg[
\cos\Big(\frac{k_bx_1}{\sqrt 6}\Big)\sin\Big(\frac{k_bx_2}{\sqrt 6}\Big)
\sin\Big(\frac{2k_bx_3}{\sqrt 6}\Big)+\\
\nonumber
\cos\Big(\frac{k_bx_2}{\sqrt 6}\Big)\sin\Big(\frac{k_bx_3}{\sqrt 6}\Big)
\sin\Big(\frac{2k_bx_1}{\sqrt 6}\Big)
+\cos\Big(\frac{k_bx_3}{\sqrt 6}\Big)\sin\Big(\frac{k_bx_1}{\sqrt 6}\Big)
\sin\Big(\frac{2k_bx_2}{\sqrt 6}\Big)
\Bigg]
 \end{eqnarray}

\end{document}